\documentclass[11pt, A4]{article}

\usepackage{graphicx}
\usepackage{epstopdf}
\usepackage{amsmath}
\usepackage{amssymb}
\usepackage{amsfonts}
\usepackage{amsthm}
\usepackage[usenames]{color}
\usepackage{array}
\usepackage{subfigure}
\definecolor{AV}{rgb}{0.65,0.0,0}
\definecolor{GC}{rgb}{0,0.0,0.65}

\usepackage[
      colorlinks=true,
      linkcolor=blue,
      urlcolor=blue,
      filecolor=blue,
      citecolor=red,
      pdfstartview=FitV,
      pdftitle={},
      pdfauthor={},
      pdfsubject={},
      pdfkeywords={},
      pdfpagemode=None,
      bookmarksopen=true
]{hyperref}

\usepackage{epsfig}

\usepackage{hyperref}

\def\cM{\mathcal{M}}

\def\cQ{\mathcal{Q}}

\def\cT{\mathcal{T}}

\def\cV{\mathcal{V}}

\def\half{\frac{1}{2}}

\def\beq{\begin{eqnarray}}
\def\eeq{\end{eqnarray}}

\def \RR{{\mathbb{R}}}

\def\mf{\mathfrak}

\def\be{\begin{equation}}
\def\ee{\end{equation}}
\def\bea{\begin{eqnarray}}
\def\eea{\end{eqnarray}}

\newcommand{\rom}[1]{\mathrm{#1}}

\def\cM{\mathcal{M}}

\def\cQ{\mathcal{Q}}

\def\cT{\mathcal{T}}

\def\cV{\mathcal{V}}

\def\mf{\mathfrak}

\def\Atwo{\omega_3}
\def\Azero{\chi_1}

\def\axone{\chi_2}
\def\axtwo{\chi_3}

\def\nn{\nonumber}
\def\half{\frac{1}{2}}

\begin{document}
\pagestyle{myheadings}
\markboth{\textsc{\small }}{%
  \textsc{\small A General Black String and its Microscopics }} \addtolength{\headsep}{4pt}


\begin{flushright}
\texttt{ULB-TH/10-23}
\end{flushright}

\begin{centering}

  \vspace{0cm}

  \textbf{\Large{A General Black String and Its Microscopics}}

  \vspace{0.8cm}

  {\large Geoffrey~Comp\`{e}re$^\natural$, Sophie~de Buyl$^\natural$, Sean Stotyn$^{ \natural, \: \sharp}$, and Amitabh Virmani$^\Diamond$ }

  \vspace{0.5cm}

\begin{minipage}{.9\textwidth}\small \it \begin{center}$^\natural$
University of California at Santa Barbara \\
CA--93106 Santa Barbara, United States \\
  {\tt gcompere@physics.ucsb.edu}$\quad$
 {\tt sdebuyl@physics.ucsb.edu}
$ \, $ \\$ \, $ \\
    $\sharp$
Department of Physics and Astronomy, University of Waterloo, \\
Waterloo, Ontario N2L 3G1, Canada \\
 {\tt smastoty@sciborg.uwaterloo.ca}\\
$ \, $ \\
$\Diamond$    Physique Th\'eorique et Math\'ematique,  \\ Universit\'e Libre de
    Bruxelles and International Solvay Institutes\\ Campus
    Plaine C.P. 231, B-1050 Bruxelles,  Belgium\\
   {\tt avirmani@ulb.ac.be}
    \end{center}
\end{minipage}

\end{centering}

\vspace{1cm}


\begin{abstract}
Using $\mathrm{G_{2(2)}}$ dualities we construct the most general
black string solution of minimal five-dimensional ungauged supergravity. The black string has five independent parameters, namely, the magnetic one-brane charge,  smeared electric zero-brane charge, boost along the string direction, energy above the BPS bound, and rotation in the transverse space. In one extremal limit it reduces to the three parameter supersymmetric string of five-dimensional minimal supergravity; in another extremal limit it reduces to the three parameter non-supersymmetric non-rotating extremal string of five-dimensional minimal supergravity. It also admits an extremal limit when it has maximal rotation in the four-dimensional transverse space. The decoupling limit of our general black string is a BTZ black hole times a two sphere. The macroscopic entropy of the string is reproduced by the Maldacena-Strominger-Witten CFT in appropriate ranges of the parameters. When the pressureless condition is imposed, our string describes the infinite radius limit of the most general class of black rings of minimal supergravity. We discuss implications our solution has for extremal and non-extremal black rings of minimal supergravity.
\end{abstract}
\vfill


\thispagestyle{empty} \newpage

\tableofcontents

\setcounter{equation}{0}

\section{Introduction}
String theory has offered significant insights in
answering many puzzles surrounding properties of black holes. Arguably, one of the greatest successes
of string theory is the statistical mechanical explanation of the Bekenstein-Hawking entropy of certain extremal and near-extremal black holes. The first example was the five-dimensional black hole studied in \cite{Strominger:1996sh}. Following \cite{Strominger:1996sh}, a number of different types of black holes have been studied and
the agreement between the statistical mechanical entropy and the Bekenstein-Hawking
entropy has been shown to hold in each case (see e.g. reviews \cite{Mathur:2005ai, David:2002wn, Peet:2000hn} and references therein).  In particular, references \cite{Callan:1996dv, Horowitz:1996fn, Breckenridge:1996sn} extended the success of \cite{Strominger:1996sh} to the five-dimensional near-extremal setting. The black holes considered in \cite{Strominger:1996sh, Callan:1996dv, Horowitz:1996fn, Breckenridge:1996sn} all have topologically spherical horizons. In 2004, Elvang, Emparan, Mateos, and Reall (EEMR) \cite{Elvang:2004rt} presented the first example of a supersymmetric solution with horizon topology $S^1\times S^2$, that is, a supersymmetric black ring. The Bekenstein-Hawking entropy of the supersymmetric black ring was reproduced from a statistical mechanical description in \cite{Cyrier:2004hj, Bena:2004tk} using the Maldacena-Strominger-Witten (MSW) CFT \cite{Maldacena:1997de}. Although, the application of MSW CFT gives an impressive match of the statistical and Bekenstein-Hawking entropies, it leaves many questions unanswered. For a critical discussion and further references see \cite{Emparan:2006mm}. It is hoped (see e.g. \cite{Emparan:2008qn}) that the matching of the statistical and Bekenstein-Hawking entropies for black rings in the thermally excited (near-extremal) setting will clarify many questions left unanswered in the supersymmetric limit. A microscopic discussion for the entropy of the yet-to-be found black ring that describes thermal excitations over the supersymmetric ring of \cite{Elvang:2004rt} is already proposed in the literature \cite{Larsen:2005qr}.

On the supergravity side, the construction of a black ring solution that describes thermal excitations over the supersymmetric black ring is a significantly involved problem.  Unlike the case of topologically spherical black holes \cite{Cvetic:1996xz}, one cannot add a sufficient number of charges on a neutral black ring by applying familiar solution generating techniques such as boosts and string dualities.  For example,  one cannot add  \cite{Elvang:2003mj}
three independent M2 charges to an otherwise neutral ring by applying boosts and string dualities. Adding the third M2 charge typically requires applying a boost along the direction of a KK-monopole in an intermediate step. Such a boost is incompatible with the identifications imposed by the KK-monopole fibration. In effect, one ends up generating pathological spacetimes.

A variant of this problem also arises in the straight string limit when one tries to add three M2 charges on top of a black string carrying three M5 charges \cite{Elvang:2004xi, Compere:2009zh}. This and related problems have resisted various attempts to construct a variety of non-extremal black string and black ring solutions \cite{Elvang:2003mj, Elvang:2004xi, Bouchareb:2007ax, Compere:2009zh,Gal'tsov:2009da}. Reference \cite{Compere:2009zh}
 shed some light on this problem from a group theoretic perspective in the simplified setting of minimal five-dimensional supergravity. There the pseudo-compact generators of the three-dimensional hidden symmetry group were associated to charging transformations.  A proposal was made to construct the solution in the straight string limit by acting with several charging transformations. The proposal however did not allow the authors of \cite{Compere:2009zh} to write the solution in a manageable form. In this paper we extend the investigations of \cite{Compere:2009zh} and present a solution of this problem in minimal supergravity in the straight string limit, i.e.,  we explicitly construct the most general black string solution of five dimensional minimal supergravity. Our black string has five independent parameters, namely, magnetic one-brane charge,  smeared electric zero-brane charge, boost along the string direction, energy above the BPS bound, and rotation in the transverse space.  In our solution one is allowed to vary all five parameters independently. Upon setting the magnetic or electric charge to zero, our solution reduces to the spinning non-extremal string solutions of \cite{Compere:2009zh, Tanabe:2008vz}. Our solution admits a four parameter family of extremal black strings corresponding to having maximal rotation in the transverse space.
We can recover the two known extremal black strings by taking additional limits.
In one limit we recover the three parameter supersymmetric string of \cite{Kim:2010bf, Shmakova:1996nz}. In another extremal limit our string solution reduces to the three parameter non-rotating non-supersymmetric string of \cite{Kim:2010bf}. When the pressureless condition is imposed, our solution describes the infinite radius limit of the most general  class of black rings of minimal supergravity.

On the microscopics side, the validity of the entropy formula in \cite{Larsen:2005qr} depends on the absence of finite ring radius corrections, which is not guaranteed \emph{a priori}.  In the straight string limit this subtlety goes away. So, it is expected that the entropy formula in \cite{Larsen:2005qr} reproduces the Bekenstein-Hawking entropy of our string in the near-extremal limit. This indeed turns out to be the case once one has made precise the connection between quantities defined near the horizon and at infinity.

An important open issue in the black ring literature has been why various black ring solutions must obey a number of bounds on various parameters in order to be smooth. For example, for the supersymmetric ring of minimal supergravity, the electric charge is bounded from below \cite{Elvang:2004rt}. The lower bound is in terms of the dipole charge and the radius of the ring \cite{Elvang:2004rt}. Curiously enough, the bound persists in the infinite radius limit. Although, we do not offer a complete understanding of these issues, our string solution does shed light on some limit of these puzzles as well. The extremal pressureless limits of our non-rotating string provide us with a working phase diagram for a class of extremal black rings of minimal supergravity. The phase diagram suggests that as the above mentioned bound on the electric charge of the supersymmetric black ring is violated, a non-supersymmetric branch of extremal black rings emerges. This non-supersymmetric branch of extremal black rings has not yet been constructed in full generality. On this branch one can continuously take the electric charge to zero, thus the non-supersymmetric branch of black rings is connected to a dipole ring of \cite{Emparan:2004wy}. The dipole ring is such that in the infinite radius limit the boost along the string completely breaks supersymmetry. Our phase diagram in addition  suggests that there is another branch of extremal black rings in minimal supergravity. This branch is also connected to a dipole ring of \cite{Emparan:2004wy}. The dipole ring on this branch is such that in the infinite radius limit the boost along the string \emph{preserves} supersymmetry. Extremal black rings on this branch are also not known in full generality.

The rest of the paper is organized as follows. In section \ref{sec:setup} we introduce our notation and conventions.  In section \ref{sec:construction} we construct the most general black string solution and analyze its physical properties and thermodynamics. First, in section \ref{charge_matrix},  we  construct a group element of G$_{2(2)}$ in order to build an appropriate `charge matrix;' for the benefit of the reader the concept of the `charge matrix' is reviewed. Next, in section \ref{sec:monster}, we act on the Kerr string with this group element to construct the most general string.  Physical properties and thermodynamics of our general solution are presented in section \ref{thermo}.  In section \ref{sec:limits} we explicitly find the extremal limits of our solution.  In \ref{extremal_limits} we show there are three distinct extremal limits, corresponding to a four-parameter non-supersymmetric string with maximal rotation in the transverse space, a three-parameter supersymmetric string, and a three-parameter non-rotating non-supersymmetric string.  In \ref{electric} and \ref{magnetic} we briefly discuss how our solution reduces to the spinning $\rom{M2}^3$ and $\rom{M5}^3$ strings respectively.  In section \ref{sec:decoupling} we match the entropy of our general solution with the prediction of \cite{Larsen:2005qr}. In section \ref{decoupling} we outline how to properly take the decoupling limit and verify that our solution in this limit is $\rom{BTZ}\times S^2$.  In section \ref{matching} we match the entropy with the CFT prediction.   We finish in section \ref{sec:discussion} with a summary and a discussion about a few implications that the work presented herein has for yet-to-be found black rings.

\setcounter{equation}{0}

\section{The set-up}
\label{sec:setup}

Since we are interested in constructing a black string solution, we can compactify minimal five-dimensional supergravity along the string direction and construct the corresponding black hole in the resulting four-dimensional theory. This way to approaching the problem turns out to be much more systematic, as there are a number of theorems \cite{Breitenlohner:1987dg} to bank on concerning black holes in the resulting four-dimensional theory. The resulting four-dimensional theory is a single modulus  $N=2,~D=4$ supergravity. It is sometimes also called the S$^3$ model (or the $t^3$ model) as its prepotential is cubic in the modulus. It can be obtained by setting S$=$T$=$U in the STU model \cite{Duff:1995sm}. In this section we review salient features of  $N=2,~D=4$ single modulus S$^3$ model. We define electromagnetic charges in this model and discuss the uplift back to minimal five-dimensional supergravity.

\subsection{$N=1,~D=5$ and $N=2,~D=4$ S$^3$  supergravities}
Let us start from the bosonic sector of eleven-dimensional supergravity
\begin{equation}
\mathcal{L}_{11} = R_{11} \star_{11}  \mathbf{1} - \half\star_{11}  \mathcal F \wedge \mathcal F +\frac{1}{6} \mathcal F \wedge \mathcal F \wedge \mathcal A, \label{11dsugra}
\end{equation}
where $\mathcal A$ is a three-form and $\mathcal F = d \mathcal A$ is its field-strength. Upon compactification on a six-torus with the ansatz
\bea
ds^2_{11} &=& ds^2_5 + dx_1^2 + dx^2_2 + \cdots + dx_6^2\\
\mathcal A &=& \frac{1}{\sqrt{3}} A \wedge (d x_1 \wedge d x_2+dx_3 \wedge dx_4 +dx_5 \wedge dx_6 )
\eea
we obtain the bosonic sector of $N=1$ (i.e., minimal) five-dimensional supergravity. The Lagrangian takes the form of Einstein-Maxwell theory with a Chern-Simons term,
\begin{equation}
\mathcal{L}_5 = R_5 \star_5  \mathbf{1} - \half\star_5  F \wedge F +\frac{1}{3\sqrt 3} F \wedge F \wedge A. \label{5dsugra}
\end{equation}
To perform the dimensional reduction to four dimensions, we assume that the extra spatial direction (denoted $z$) is compact and is a Killing vector in the five-dimensional spacetime. Using the standard Kaluza-Klein ansatz to yield a four-dimensional Lagrangian in Einstein frame we write the five-dimensional metric as
\begin{align}
ds^2_5 &= e^{\frac{1}{\sqrt 3}\phi_1} ds^2_4 + e^{-\frac{2}{\sqrt 3}\phi_1}( dz+ A_1)^2, \label{eqn:metric5to4d} \\
A &= A_2 + \axone  dz. \label{pot5to4d}
\end{align}
From this
ansatz one finds (see for example \cite{Cremmer:1999du}) that the resulting four-dimensional Lagrangian takes the form
\begin{align}
\mathcal{L}_4 &= R_4 \star_4 \mathbf{1} - \frac{1}{2} \star_4 d \phi_1 \wedge  d\phi_1 - \frac{1}{2} e^{\frac{2}{\sqrt{3}} \phi_1} \star_4  d\chi_2 \wedge  d\chi_2  - \frac{1}{2} e^{-\sqrt{3}\phi_1}\star_4 F_1 \wedge F_1 \nn \\
&\quad\, \: - \: \frac{1}{2} e^{-\frac{1}{\sqrt{3}}\phi_1} \star_4 F_2 \wedge F_2 + \frac{1}{\sqrt{3}} \, \chi_2 \,  dA_2 \wedge  dA_2,
\label{lagrangian}
\end{align}
where
\begin{align}
\label{eqn:4dfieldstrengths}
F_1 &= d A_1, &
F_2 &= d A_2 - d\chi_2 \wedge A_1.
\end{align}
The scalars $\phi_1$ and $\chi_2$ parameterize an $\mathrm{SL}(2, \mathbb{R})/\mathrm{U}(1)$ coset. The four electromagnetic charges corresponding to the two vectors are defined as follows.
The equations of motion for the potentials $A_1$ and $A_2$ are
\begin{equation}\label{eqn:eom4A1}
d\big( e^{-\sqrt{3} \phi_1} \star_4 F_1)+e^{-\frac{1}{\sqrt{3}} \phi_1} \star_4 F_2\wedge d\chi_2 =0,
\end{equation}
and
\begin{equation}\label{eqn:eom4A2}
d\beta_2 \equiv d\big( e^{-\frac{1}{\sqrt{3}} \phi_1} \star_4 F_2 - \frac{2}{\sqrt{3}} \chi_2 d A_2 \big) = 0.
\end{equation}
Using \eqref{eqn:eom4A2}, one can rewrite \eqref{eqn:eom4A1} as the closure of a form $\beta_1$,
\begin{equation}\label{eqn:eom4A1-1}
d\beta_1 \equiv d\big( e^{-\sqrt{3} \phi_1} \star_4 F_1+e^{-\frac{1}{\sqrt{3}} \phi_1} \star_4 F_2 \chi_2 - \frac{1}{\sqrt{3}} d A_2 \chi_2^2\big)=0.
\end{equation}
We then use the closed forms $\beta_1$ and $\beta_2$ given by \eqref{eqn:eom4A1-1} and \eqref{eqn:eom4A2} to define conserved electric charges in asymptotically flat four-dimensional spacetimes as integrals over a two-sphere at spatial infinity $S^2_\infty$
\begin{align}
Q_{1} &= \frac{1}{4 \pi}  \int_{S^2_{\infty}} \beta_1,&
Q_{2}  &= \frac{1}{4 \pi \sqrt{3}}  \int_{S^2_{\infty}} \beta_2.
\end{align}
In a similar fashion, from the Bianchi identities
\begin{equation}
d F_1 = 0, \qquad
d( F_2 + d \chi_2 \wedge A_1 ) = 0,
\end{equation}
for $A_1$ and $A_2$ we define the magnetic charges
\begin{align}
P_1 &=- \frac{1}{4 \pi}  \int_{S^2_{\infty}} F_1, &
\label{P1P2}
P_2 &=  \frac{1}{4 \pi \sqrt{3}}  \int_{S^2_{\infty}} F_2 + d\chi_2 \wedge A_1.
\end{align}
Note the minus sign in the definition of $P_1$. We work with sign conventions in which the static extremal black hole carrying positive $P_1$ and $Q_2$ charges
is supersymmetric. From the M-theory point of view the electromagnetic charges correspond to the brane charges as indicated in Table \ref{tablecharges} :\footnote{For more details on the brane interpretation see the brane intersection tables in \cite{Compere:2009zh}.}
\vskip 0.2cm
\begin{table}
\begin{center}
\begin{tabular}{|c|c|}\hline
$Q_1$ & Kaluza-Klein momentum (P) along the M-theory circle  \\ \hline
$P_1$ & Kaluza Klein-monopole (KKM) along the M-theory circle \\ \hline
$Q_2$ & M2$^3$ (three equal M2 charges)  \\ \hline
$P_2$ &  M5$^3$ (three equal M5 charges) \\ \hline
\end{tabular}
\caption{Four dimensional
electric and magnetic charges interpreted in M theory.
\label{tablecharges}}
\end{center}
\end{table}
\vskip 0.2 cm

\noindent Thus, the $(Q_1, P_2)$ system with $Q_1,\,P_2 > 0$ corresponds to M5$^3$ -- P, which is BPS.
The $(Q_2, P_1)$ system with $Q_2,\,P_1 > 0$ corresponds to
M2$^3$ -- KKM, which
is also BPS.

We now define two scalar charges for $\phi_1$ and $\chi_2$ as the radial derivatives of these fields at spatial infinity. In this paper, we impose the condition that  both scalars vanish at spatial infinity. This is a consequence (accompanied with a natural gauge choice) of the five-dimensional Kaluza-Klein boundary conditions. With this condition imposed, the scalar charges can simply be defined as
\begin{align}
\Sigma_s &= \lim_{r\to\infty}\frac{r\,\phi_{1}(r)}{\sqrt3},&
\Xi&= \lim_{r\to\infty}\frac{r\,\chi_{2}(r)}{\sqrt3}.
\end{align}
For further details we refer the reader to \cite{Kim:2010bf}.

\subsection{Reduction on time}
\label{sec:reductionontime}
In this paper we exclusively work with stationary spacetimes.  Therefore, we also assume the existence of a timelike Killing vector $\partial_t$ commuting with $\partial_z$. Now we can reduce the theory to three dimensions over this timelike Killing vector. The standard Kaluza-Klein ansatz for this reduction is
\begin{align}
\label{eqn:metric4dto3d}
ds^2_4  &= e^{\phi_2}ds^2_3  - e^{-\phi_2}(dt + \Atwo)^2, \\
\label{eqn:metricMaxwellto3d}
A_1 & =  B_1 + \Azero dt, \\
\label{eqn:Maxwellto3d}
A_2 & =  B_2+ \axtwo dt.
\end{align}
From this reduction we end up with three-dimensional Euclidean gravity coupled to five scalars and three one-forms. The one-forms $B_1, B_2,$ and $\omega_3$ can be dualized into scalars. Scalars dual to the one-forms $B_1$, $B_2$, and $\omega_3$ in the notation of \cite{Compere:2009zh} are denoted by $\chi_5$, $\chi_4$, and $\chi_6$, respectively\footnote{Note that in \cite{Compere:2009zh} the reduction was performed first along time and then along $z$. The two cosets obtained by either order of the reduction are related by non-trivial field redefinitions.  It turns out that reducing first over space and then time is far more convenient, and this is the approach we follow in this paper.}. Upon dualization of the one-forms into scalars the Lagrangian in three dimensions becomes Euclidean gravity coupled to eight scalars. The scalars parameterize the pseudo-Riemannian coset $\mathrm{G}_{2(2)}/\tilde{K}$, with
\begin{align}
\tilde{K} = \mathrm{SO}_0(2,2) \cong  (\mathrm{SL}(2, \RR) \times \mathrm{SL}(2, \RR))/\mathbb{Z}_2.
\end{align}
Let $\cV$ be the coset representative for the coset $\mathrm{G}_{2(2)}/\tilde{K}$, then a matrix $\cM$ is defined as
\beq
\cM := (\cV^{\sharp})  \cV, \label{M}
\eeq
where $\sharp$ stands for the generalized transposition. It is
defined on the generators of the Lie algebra $\mf{g}_{2(2)}$ by
\beq
x^\sharp := -  \tau (x),\, \qquad  \forall \, x \in \mf{g}_{2(2)},
\label{generalizedtransposition}
\eeq
where the involution $\tau$ defines the coset. Under the group action, the matrix $\mathcal{M}$ transforms as
\beq
\cM \rightarrow g^\sharp \, \cM \, g \hspace{1cm} \mbox{when} \hspace{1cm}\cV \rightarrow  \cV \, g, \hspace{1cm} g \in \mathrm{G}_{2(2)}.
\eeq
For further details and explicit formulae on the coset construction we refer the reader to \cite{Compere:2009zh, Kim:2010bf}.
See also \cite{Bouchareb:2007ax, Giusto:2007fx, Gal'tsov:2008nz, Figueras:2009mc}.

Using the reduction ansatz (\ref{eqn:metric4dto3d}) we can calculate the mass and NUT charge explicitly in terms of the three-dimensional fields. Following \cite{Bossard:2008sw}, we define the Komar mass and NUT charge as
\begin{align}
M &= \frac{1}{8 \pi G_4} \int_{S^2_{\infty}} \star_4 K, & N &=  \frac{1}{8 \pi G_4} \int_{S^2_{\infty}}  K,
\end{align}
where $K=dg$, $g=g_{\mu\nu} \kappa^\mu dx^\nu$, and  $\kappa = \partial_t$. From the metric ansatz (\ref{eqn:metric4dto3d}) we find
\begin{equation}
K = \partial_{\nu} g_{t \mu} dx^{\nu} \wedge dx^{\mu},
\end{equation}
which yields
\begin{equation}
M = -\frac{1}{8 \pi G_4} \int_{S^2_{\infty}} \partial_r e^{-\phi_2} \,\star_4 (dr \wedge dt),
\end{equation}
where, in our conventions, $\star_4 ( d r \wedge d t) = - e^{\phi_2}r^2 \sin{\theta}  d \theta \wedge d \phi$.
From asymptotic flatness we have
\begin{equation}
\phi_2 (r)= \frac{2 G_4 M}{r}+ \mathcal{O}\left(\frac{1}{r^2}\right).
\end{equation}
(For a detailed discussion of boundary conditions we refer the reader to \cite{Bossard:2008sw}.) Thus,
\begin{align}
G_4 M = \lim_{r\to\infty}\frac{r\phi_{2}(r)}{2}.
\end{align}

To summarize, the ansatz
\begin{align}
ds^{2}_{5} &=
e^{\frac{1}{\sqrt3} \phi_{1}+\phi_{2}}ds^{2}_{3}
-e^{\frac{1}{\sqrt3} \phi_{1}-\phi_{2}}  (dt+\omega_3)^2
+e^{-\frac{2}{\sqrt3} \phi_{1}} ( dz +B_1 + \chi_1 dt)^{2},\label{eqn:canonicalform}
\\
\label{eqn:canonicalMaxwell}
A&=B_2+\chi_{3}dt + \chi_{2}dz,
\end{align}
describes stationary solutions of the S$^3$ model uplifted to five dimensions. It follows from four-dimensional asymptotic flatness that the electric and magnetic charges defined above can  also be expressed in terms of asymptotic values of the scalars as
\begin{align}
Q_1&=   \lim_{r \to \infty} \,r \chi_1(r),& 
Q_2&=    \lim_{r \to \infty} \, \frac{r \chi_3(r)}{\sqrt{3}},\nn\\
P_1&= \lim_{r \to \infty} \,r \,\chi_5(r),& 
P_2&= -\lim_{r \to \infty} \, \frac{r\, \chi_4(r)}{\sqrt{3}}.
\end{align}
Similarly, the NUT charge can also be expressed as \be
N = -\lim_{r\to\infty}\frac{r\chi_{6}(r)}{2G_4}.
\ee

\subsection{Five dimensional charges}
For our black string the spatial part of the metric at infinity is $\mathbb{R}^{3} \times S^1$. For asymptotically Kaluza-Klein spacetimes,
the five-dimensional ADM mass $M_5$, linear momentum $P_z$ along the string direction, and ADM tension $\mathcal T$ are defined as (see e.g. \cite{Harmark:2004ch} and section 2.1 of \cite{Myers:1999psa})
\bea
M_5 &=& 2 \pi R \: T_{tt}  = \frac{\pi R}{2G_5}(2c_{tt}-c_{zz}),\\
\mathcal T &=& -T_{zz} = \frac{1}{4G_5}( c_{tt}-2c_{zz}),\\
P_z &=& 2\pi R \: T_{tz} = \frac{\pi R}{2G_5} c_{tz},
\eea
where $c_{tt}, c_{tz},$ and $c_{zz}$ are the leading corrections of the metric at infinity
\bea
g_{tt} \simeq - 1 + \frac{c_{tt}}{r}, \qquad g_{zz} \simeq  1 + \frac{c_{zz}}{r}, \qquad g_{tz} \simeq \frac{c_{tz}}{r}.
\eea
One can re-express these charges in terms of four-dimensional charges as
\bea
M_5 = M,\qquad
\mathcal T  = \frac{1}{8 \pi R G_4}( 3\Sigma_s +2 G_4 M ),\qquad
P_z = \frac{Q_1}{4G_4},
\eea
where we have used $G_5 = (2\pi R) G_4$.
The five-dimensional electric charge in geometrized units for asymptotically Kaluza-Klein spaces is
\bea
Q_\rom{E} \equiv \frac{1}{16\pi G_5}\int_{S^2 \times \mathbb R}\star_{5}F =\frac{\sqrt{3}}{4 G_4}Q_2,
\eea
where the Chern-Simons term has been set to zero using the boundary conditions at infinity. The magnetic one-brane charge is
\be
Q_\rom{M} \equiv \frac{1}{4\pi} \int_{S^2} F = \sqrt{3} P_2.
\ee

\setcounter{equation}{0}

\section{Construction and physical properties of the general string}
\label{sec:construction}

In this section we present a construction and physical properties of our general string solution. To construct the solution, we proceed in two steps. In the first step (section \ref{charge_matrix}), we construct an appropriate `charge matrix'. In the second step (section \ref{sec:monster})
we use the group element $k \in \tilde K$ used to generate the charge matrix
to construct our general string. This two step decomposition of the problem is somewhat artificial, but it helps in disentangling spacetime physics from group theory. For the benefit of reader we also briefly review the concept of charge matrix in section \ref{charge_matrix}. Physical properties and thermodynamics of our solution are presented in section \ref{thermo}.

\subsection{Charge matrix}
\label{charge_matrix}
Since the work of Breitenlohner, Maison, and Gibbons \cite{Breitenlohner:1987dg} it is known that single center spherically symmetric black holes for a wide class of four-dimensional gravity theories correspond to geodesic segments on coset manifolds $G/{\tilde K}$.  A geodesic on the coset manifold is completely specified by its starting point $p \in G/{\tilde K}$ and its velocity at $p$. The velocity at the point $p$ is the conserved Noether charge $\mathcal{Q} \in \mathfrak{g}$ taking values in the Lie algebra $\mathfrak{g}$ of $G$. From the four-dimensional spacetime point of view, the starting position $p$ of the geodesic is associated to the values of the moduli at spatial infinity, and the velocity $\mathcal{Q}$ at the point $p$ is associated to the four-dimensional conserved charges.

The action of $G$ on a given solution is such that it acts both on the position and the velocity of the corresponding geodesic.  The subgroup of $G$ that keeps the starting point $p$ fixed is $\tilde K$. The subgroup $\tilde K$ thus generates the full set of transformations of the conserved charge $\cQ$. It was shown in \cite{Breitenlohner:1987dg} that using $\tilde K$ one can generate the full class of single-center non-extremal spherical symmetric black holes starting from the Schwarzschild black hole.  Reference \cite{Breitenlohner:1987dg} also showed that using $\tilde K$ one can generate the full class of single-center non-extremal \emph{rotating} black holes starting from the Kerr black hole. In other words, all single-center non-extremal black holes with requisite symmetry lie in a single $\tilde K$-orbit containing the Kerr black hole.

If we assume asymptotic flatness and that the geodesic corresponding to a given black hole starts at the identity coset, then the matrix $\cM$ admits an expansion in powers of the radial coordinate \begin{equation}
\cM = 1 + \frac{ \mathcal{Q}}{r} + \ldots,
\end{equation}
where $\mathcal{Q} \in \tilde{\mathfrak{p}}$, where
 $ \tilde{\mathfrak{p}}$ is the complement of
the Lie algebra of $\tilde K$ in $\mf{g}_{2(2)}$ with respect to the Killing form.
 The matrix $\cQ$ is also called the charge matrix.
In \cite{Bossard:2009at} it was  observed, following \cite{Breitenlohner:1987dg}, that the charge matrix $\mathcal{Q}$ for any four-dimensional asymptotically flat non-extremal axisymmetric solution satisfies
\begin{equation}
\mathcal{Q}^3 - \frac{\mathrm{tr}(\mathcal{Q}^2)}{4} \mathcal{Q} = 0. \label{characterstic}
\end{equation}
 The information about the rotation in the four-dimensional spacetime is not contained in the charge matrix. Rotation enters at the next to the leading term in the expansion of the matrix $\cM$. The structure of the solutions is however such that from the charge matrix  and the Kerr geometry one can uniquely construct the full spacetime solution \cite{Breitenlohner:1987dg, Bossard:2009at}.

In terms of the electromagnetic and scalar charges defined above,  the most general charge matrix for the S$^3$ model is written as  \cite{Kim:2010bf}
\be
\mathcal{Q}=-4 G_4 Mh_1+(-\Sigma_s-2G_4 M)h_2-Q_1 p_1+\Xi p_2-Q_2 p_3+\tfrac12 P_2 p_4+\tfrac16 P_1p_5+\tfrac13 G_4 N p_6,\label{generalchargematrix}
\ee
where we again follow the conventions for the generators given in  \cite{Kim:2010bf,Compere:2009zh}.
The Schwarzschild solution simply corresponds to
\be
\mathcal{Q}_\rom{Schw}=-2  m (2 h_1+h_2),
\ee
where $m= G_4 M$, and $M$ is the ADM mass of the Schwarzschild solution. To construct the solution we are after, we proceed in two steps. In the first step, we construct an appropriate charge matrix by acting  on $\mathcal{Q}_\rom{Schw}$ with an element $k \in \tilde K$
  \be
   -2 m   k^{-1} \cdot (2 h_1 + h_2) \cdot k.
  \ee
Once we have the appropriate charge matrix, acting with the corresponding $k$ on $\cM_\rom{Schw}$ one can construct the non-rotating string with appropriate charges.   In the second step we use the \emph{same} $k$ and act with it on $\cM_\rom{Kerr}$.   The first problem can be solved by systematically studying the action of various generators of $\tilde K$ on the general charge matrix \eqref{generalchargematrix}. On the technical side one still has to overcome three main obstacles: $(i)$ when acting with $\tilde K$ the resulting charge matrix is in general  parameterized in an unilluminating manner, so one has to choose parameters judiciously, $(ii)$ in order to obtain
regular black holes, one has to force the NUT charge to vanish $N=0$, $(iii)$ in order for the solution to correspond to a black string of five-dimensional minimal supergravity one also has to force the KK-monopole charge to vanish $P_1=0$. All of this can be achieved by the following element of the group $\tilde K$,
\bea
k = e^{-\frac{1}{2}\log\alpha \, k_3}e^{{1\over 2} \beta k_4}e^{- \beta k_1}e^{-\frac{1}{2}\log(\tilde \delta/\alpha) k_3}e^{-\pi k_2}e^{-\gamma k_1} \label{generatork}
\eea
with
\be
\alpha^3 = \frac{\tilde \delta^2 (3 + \tilde \delta)}{(1 + 3 \tilde \delta)}.
\ee
We have found this group element by a process involving
a considerable amount of trial and error, followed
by an explicit verification of the physical conditions mentioned above. The final charge matrix indeed satisfies \eqref{characterstic}.
The non-extremality parameter is simply
\be
c^2 := \frac{1}{4}\mathrm{tr}(\mathcal{Q}^2) = 4  m^2.
\ee
There are only three independent parameters in the group element $k$, namely, $\beta, \tilde \delta$ and $\gamma$. Choosing $\tilde \delta = \exp(2 \delta)$ we find that the charge matrix is parameterized as:
\bea
G_4 M &=& m \left( -\frac{2 c_\beta c_\gamma s_\beta s_\gamma}{\sqrt{1 + 3 c_\delta^2}}+ {1\over 2}(c_\beta^2 + s_\beta^2) (c_\delta^2 (1 +  c_\gamma^2) + s_\delta^2 (s_\gamma^2 + c_\gamma^2))\right), \label{charge1}\\
Q_1 &=& 2 m\left( \frac{2 c_\beta s_\beta (c_\gamma^2 + s_\gamma^2)}{\sqrt{1 + 3 c_\delta^2}}- c_\gamma s_\gamma (c_\beta^2 + s_\beta^2)(1 + 3 s_\delta^2)\right),\\
Q_2 &=& 2m c_\delta s_\delta \left(  \frac{2 s_\gamma s_\beta c_\beta}{\sqrt{1 + 3 c_\delta^2}} + c_\gamma (c_\beta^2 + s_\beta^2)\right),\\
P_2 &=&
\frac{4m c_\beta   s_\beta c_\delta^2 }{\sqrt{1 + 3 c_\delta^2}},
\\
\Sigma_s &=&\frac{4m c_\beta c_\gamma s_\beta s_\gamma}{\sqrt{1 + 3 c_\delta^2}} - m(c_\beta^2 + s_\beta^2) (s_\gamma^2 c_\delta^2 + s_\gamma^2 s_\delta^2 + c_\gamma^2 s_\delta^2), \\
\Xi &=& - 2m c_\delta  s_\delta \left( \frac{2  c_\beta c_\gamma  s_\beta }{\sqrt{1 + 3 c_\delta^2}} + (c_\beta^2 + s_\beta^2)s_\gamma\right), \\
N &= & 0, \qquad P_1 = 0,
\label{charge8}
\eea
where to avoid notational clutter we have defined $c_\delta = \cosh \delta,~s_\delta = \sinh \delta,~c_\gamma = \cosh \gamma,~s_\gamma = \sinh \gamma,~c_\beta = \cosh \beta,~s_\beta = \sinh \beta$. Expressions \eqref{charge1}--\eqref{charge8} are an important result of this paper.

\subsection{The general black string solution}
\label{sec:monster}
To explicitly construct the solution we are after we now act with $k$ on the matrix $\mathcal M_{\rom{Kerr}}$ and read of the new scalars from it. We then reconstruct the five-dimensional solution. Our seed solution is the direct product of the Kerr geometry times a line along the $z$ direction. This metric can be written in the convenient form
\be
ds^2 = -\frac{\Delta_2}{\Sigma}(dt + \omega^\rom{seed}_\phi d\phi)^2 +\Sigma \, ds_\rom{base}^2 + dz^2, \hspace{1cm} \epsilon_{r x \phi tz} = - \Sigma,\label{Kerr}
\ee
where
\bea
ds^2_\rom{base} &=& \left( \frac{dr^2}{\Delta}+\frac{dx^2}{1-x^2} \right) + \frac{\Delta (1-x^2)}{\Delta_2}d\phi^2,
\eea
\be
\omega^\rom{seed}_\phi = -2amr\frac{1-x^2}{\Delta_2},
\ee
and
\bea
\Delta &=& r^2 - 2m r+a^2,\qquad \Delta_2 = r^2 - 2 m r +a^2 x^2 ,\qquad \Sigma = r^2 + a^2 x^2.
\eea
For calculational convenience we use $x = \cos \theta$ as one of the coordinates, instead of the polar coordinate $\theta$.
Upon reducing this seed solution to three dimensions, constructing the matrix $\mathcal M_\rom{Kerr}$, acting with the group element $k$ and lifting back to five dimensions, we find the most general asymptotically  Kaluza-Klein black string solution of five-dimensional minimal supergravity. The solution is endowed with five independent parameters: the magnetic one-brane charge (this corresponds to the parameter $\beta$), the  smeared  electric zero-brane charge (this corresponds to the parameter $\delta$), boost along the string direction (roughly speaking this corresponds to the parameter $\gamma$, see additional comments in section \ref{thermo}), energy above the BPS bound (this corresponds to the parameter $m$), and rotation in the transverse space (this corresponds to the parameter $a$).
The line element and gauge potential for $\gamma=0$ can be written as
\begin{align}
ds^2 &=& (F_1+\Delta_2) \left[ -  \frac{\Delta_2}{\xi} (dt + \omega_{\phi} d \phi)^2 + ds^2_\rom{base}\right] +\frac{\xi}{(F_1+\Delta_2)^2} \left( dz + \hat A_{t} dt + \hat A_{\phi} d \phi \right)^2, \label{monster}
\end{align}
\bea
A_t &=& -\sqrt{3} \frac{F_3}{(F_1+\Delta_2)},\qquad A_z = -\sqrt{3}\frac{F_2}{(F_1+\Delta_2)},\\
 A_\phi &=& \frac{2\sqrt{3}m c_{\delta}^2}{\Delta_2}\left( -\frac{2s_{\beta}c_{\beta}\Delta x}{\sqrt{1+3c_{\delta}^2}}+a(1-x^2)s_{\delta}\left(2ms_{\beta}^2\left(\frac{6c_{\beta}^2 c_{\delta}^2} {1+3c_{\delta}^2}-1\right)+r(c_{\beta}^2+s_{\beta}^2)\right)\right)\nonumber\\ &&-\sqrt{3}\bar A_\phi \frac{F_2}{(F_1+\Delta_2)}-\sqrt{3}\omega_\phi \frac{F_3}{(F_1+\Delta_2)},
\label{monster_metric}
\eea
where we have defined the following functions
\bea
\hat A_\phi &=&  \bar A_\phi + \frac{k}{\xi}\omega_\phi, \qquad \qquad
\hat A_t  =\frac{k}{\xi},\\
\omega_\phi &=& - 2am \frac{1-x^2}{\Delta_2}c_{\delta}^3\left(2ms_{\beta}^2\left(c_{\beta}^2+s_{\beta}^2-\frac{4c_{\beta}^2}{1+3c_{\delta}^2}\right)+r(c_{\beta}^2+s_{\beta}^2)\right),\\
\bar A_\phi &=& \frac{4am(1-x^2)s_{\beta}c_{\beta}c_{\delta}^3(r+2s_{\beta}^2m)}{\Delta_2\sqrt{1+3c_{\delta}^2}},\\
\xi &=& (F_4+\Delta_2) (F_1+\Delta_2) -F_2^2,\qquad \qquad k = F_5(F_1+\Delta_2) -F_2 F_3.
\eea
The remaining five functions $F_1,F_2,F_3,F_4,F_5$ are linear in $r$ and $x$ and are given by
\bea
F_1 &=& 2mc_{\delta}^2\left(2ms_{\beta}^2\left(s_{\beta}^2+\frac{s_{\delta}^2c_{\beta}^2}{1+3c_{\delta}^2}\right)+r(c_{\beta}^2+s_{\beta}^2)+\frac{2axs_{\delta}s_{\beta}c_{\beta}}{\sqrt{1+3c_{\delta}^2}}\right),\nn \\
F_2 &=& 2mc_{\delta}s_{\delta}\left(\frac{2ms_{\beta}c_{\beta}}{\sqrt{1+3c_{\delta}^2}}\left(-1+(c_{\beta}^2+s_{\beta}^2)c_{\delta}^2\right)+\frac{2rs_{\beta}c_{\beta}}{\sqrt{1+3c_{\delta}^2}}+axs_{\delta}(c_{\beta}^2+s_{\beta}^2)\right),\nn \\
F_3 &=& 2mc_{\delta}\left(2ms_{\delta}s_{\beta}^2\left(\frac{c_{\beta}^2(1+c_{\delta}^2)}{1+3c_{\delta}^2}-s_{\beta}^2\right)-r(c_{\beta}^2+s_{\beta}^2)s_{\delta}+\frac{2axs_{\beta}c_{\beta}}{\sqrt{1+3c_{\delta}^2}}\right),\nn \\
F_4 &=& 2m\left(2m\left((c_{\beta}^2s_{\delta}^2+s_{\beta}^2c_{\delta}^2)^2+\frac{s_{\delta}^2s_{\beta}^2c_{\beta}^2}{1+3c_{\delta}^2}\right)+r(c_{\beta}^2+s_{\beta}^2)(c_{\delta}^2+s_{\delta}^2)-\frac{2axs_{\delta}s_{\beta}c_{\beta}}{\sqrt{1+3c_{\delta}^2}}\right),\nn \\
F_5 &=& 2m\left(\frac{2ms_{\beta}c_{\beta}}{\sqrt{1+3c_{\delta}^2}}\left(-1+(c_{\beta}^2+s_{\beta}^2)c_{\delta}^4\right)+\frac{2rs_{\beta}c_{\beta}}{\sqrt{1+3c_{\delta}^2}}-axs_{\delta}^3(c_{\beta}^2+s_{\beta}^2)\right).
\eea
The parameter $\gamma$ can be added by simply applying a boost to this solution
\be
t \rightarrow t \cosh \gamma - z \sinh \gamma, \qquad z \rightarrow - t \sinh \gamma + z \cosh \gamma.
\ee
After we obtained the solution, an important and difficult task was to rewrite it in the form presented above.  The range of the parameters $\beta$ and $\delta$ can be restricted to non-negative values without loss of generality. The solution with parameters $(m,a,\beta,\delta,\gamma)$ are related to the solution with parameters $(m,-a,-\beta,\delta,-\gamma)$ with the change of coordinates $\phi \rightarrow -\phi$, $z \rightarrow -z$ and to the solutions of parameters $(m,a,\beta,-\delta,\gamma)$ with the change of coordinates $\phi \rightarrow -\phi$, $z \rightarrow -z$, $t \rightarrow -t$, $x \rightarrow -x$.

\subsection{Thermodynamics}\label{thermo}
We now turn our attention to the thermodynamics of our general black string solution in the boosted frame, i.e., ADM momentum along the string non-zero, $P_z \neq 0$.
Expressions for the asymptotic charges and horizon quantities simplify substantially if we introduce the three following functions
\bea
G[X,Y] &=& \frac{-2s_{\beta}c_{\beta}XY}{\sqrt{1+3c_{\delta}^2}}+\frac{1}{4}(c_{\beta}^2+s_{\beta}^2)(X^2-2Y^2+3X^2(c_{\delta}^2+s_{\delta}^2)),\\
H[X,Y] &=& c_{\delta}\left(\frac{2s_{\beta}c_{\beta}Y}{\sqrt{1+3c_{\delta}^2}}+(c_{\beta}^2+s_{\beta}^2)X\right),\\
F[X,Y]&=&2mc_{\delta}^3\left(\left(c_{\beta}^4r_++s_{\beta}^4r_-+\frac{6ms_{\beta}^2c_{\beta}^2s_{\delta}^2}{1+3c_{\delta}^2} \right)X - \frac{2s_{\beta}c_{\beta}(c_{\beta}^2r_++s_{\beta}^2r_-)}{\sqrt{1+3c_{\delta}^2}}Y\right),
\eea
where
\bea
r_+ = m  +\sqrt{m^2-a^2}, \hspace{1cm} r_- =  m  -\sqrt{m^2-a^2},
\eea
are the outer and inner horizon radii.
In terms of these functions, the ADM mass, ADM tension, linear momentum along the string direction, angular momentum in the transverse space, horizon area, and angular and linear velocities at the outer horizon are found to be
\bea
M_5 &=& \frac{2\pi Rm}{G_5}G[c_\gamma,s_\gamma],\qquad \mathcal T =- \frac{m}{G_5}G[s_\gamma,c_\gamma],\label{eq:T2}\\
P_z &=& \frac{m\pi R}{G_5} \left(-\frac{2s_{\beta}c_{\beta}(s_{\gamma}^2+c_{\gamma}^2)}{\sqrt{1+3c_{\delta}^2}}+c_{\gamma}s_{\gamma}(c_{\beta}^2+s_{\beta}^2)(1+3s_{\delta}^2)\right),\\
J_\phi &=&  \frac{2\pi R m a c_{\delta}^2}{G_5}H[c_\gamma,-s_\gamma],\\
 A_{\rom{H}}&=&8\pi^2R \, F[c_\gamma,s_\gamma],\label{eq:A}\\
\Omega_{\phi} &=& \frac{a}{F[c_\gamma,s_\gamma]},\qquad v_z=\frac{F[s_\gamma,c_\gamma]}{F[c_\gamma,s_\gamma]}.
\eea
The temperature can be calculated from the surface gravity, which results in
\be
T=\frac{r_+ - r_-}{4 \pi F[c_\gamma,s_\gamma]}.\label{eq:T}
\ee
This, of course, vanishes for extremal solutions where $a=\pm m$.  When $a=0$ and the other parameters are finite, the function $F[X,Y]$
scales as $F[X,Y]\propto m^2$ and so the temperature scales in the usual Schwarzschild way as $T\propto m^{-1}$.

The electric and magnetic charges are given by 
\bea
Q_\rom{E} &=& \frac{\sqrt{3}m\pi Rs_{\delta}}{G_5}H[c_\gamma,s_\gamma],\qquad  Q_\rom{M} = \frac{4\sqrt{3}ms_{\beta}c_{\beta}c_{\delta}^2}{\sqrt{1+3c_{\delta}^2}}.
\eea
To calculate the chemical potentials associated with these two charges we follow \cite{Copsey:2005se}. To this end, we start by looking at the gauge field carefully.  When the magnetic charge is non-zero, the gauge field asymptotically behaves as
\be
A_\phi\sim Q_\rom{M} x +{\mathcal O}(r^{-1}), \quad\quad A_z\sim{\mathcal O}(r^{-1}), \quad\quad A_t\sim{\mathcal O}(r^{-1}).
\ee
The angular component is not well-defined simultaneously on both the north and the south pole, so we instead introduce regular gauge potentials on two patches that smoothly cover the poles.  We denote the boundary between the two patches by $E$, i.e., $E$ is the surface of constant $t$ and $\theta=\pi/2$.  The angular parts of the gauge potentials then satisfy the boundary conditions
\bea
A_\phi^\rom{North}\sim Q_\rom{M}(x-1)+{\mathcal O}(r^{-1}),\\
A_\phi^\rom{South}\sim Q_\rom{M}(x+1)+{\mathcal O}(r^{-1}).
\eea
We ultimately require a quantity of the form $\xi^\mu A_\mu$ to be continuous across $E$, where $\xi^\mu=\Omega_\phi \partial_\phi+v_z \partial_z +\partial_t$ is the generator of the horizon.  To this end we introduce the constant $\Lambda$ such that
\be
\Lambda^\rom{North}=\Omega_\phi Q_\rom{M},\quad\quad \Lambda^\rom{South}=-\Omega_\phi Q_\rom{M}
\ee
and we see that $\xi^\mu A_\mu+\Lambda$ is indeed continuous across the boundary $E$.  We are now suited to calculate the electric and magnetic potentials. The electric potential that appears in the first law is given by 
\begin{equation}
\Phi_{E}=-(\xi^\mu A_\mu+\Lambda)\Big{|}_{r_+}.
\end{equation}
It yields the result
\be
\Phi_\rom{E}=\frac{2\sqrt{3}mc_{\delta}^2s_{\delta}}{F[c_{\gamma},s_{\gamma}]}\left(r_+c_{\beta}^2\left(\frac{6c_{\delta}^2s_{\beta}^2}{1+3c_{\delta}^2}+1\right)+r_-s_{\beta}^2\left(\frac{6c_{\delta}^2c_{\beta}^2}{1+3c_{\delta}^2}-1\right)\right).
\ee

In reference \cite{Copsey:2005se} a general expression for the magnetic potential was given in the Hamiltonian form, which can be expressed in the Lagrangian form as \cite{Compere:2009zh}
\be
\Phi_\rom{M}=\frac1{8\pi}\int_E(d^3x)_{\mu\nu}(2\xi^\mu F^{\alpha\nu}\partial_\alpha\phi-\Omega_\phi F^{\mu\nu})+\frac1{8\sqrt{3}\pi}\int_E F_{\alpha\beta}\partial_\gamma\phi(A_\rho\xi^\rho+\Lambda)dx^\alpha\wedge dx^\beta\wedge dx^\gamma
\ee
where $(d^3x)_{\mu\nu}=\frac1{2!3!}\epsilon_{\mu\nu\alpha\beta\gamma}dx^\alpha\wedge dx^\beta\wedge dx^\gamma$.  Although one could in principle use this expression to calculate the magnetic potential for our solution, in practice such a computation is extremely involved and not very illuminating. Instead we use the Smarr relation to guess the magnetic potential in terms of the other physical quantities. The Smarr relation takes the form
\be
M=\frac32\left(\frac1{4G_5}T A +\Omega_\phi J_\phi+v_z P_z\right)+\frac12{\mathcal T}(2\pi R)+\Phi_\rom{E} Q_\rom{E}+\frac12\Phi_\rom{M} Q_\rom{M}.
\ee
Upon solving for $\Phi_\rom{M}$, it is a straightforward exercise, but a non-trivial check, to verify that the first law
\be
dM=\frac1{4G_5}TdA+\Omega_\phi dJ_\phi+ \frac{v_z}{R} d\left(P_z R\right)+2\pi {\mathcal T}dR+\Phi_\rom{E} d Q_\rom{E}+\Phi_\rom{M} dQ_\rom{M},
\ee
holds. Stated differently, one can assume that the first law holds and derive the magnetic potential $\Phi_\rom{M}$ from the first law. The fact that it is possible to do so, and it is a non-trivial check of the correctness of our expressions, is because the first law involves six different equations, one for the variation of each parameter $(m,a,\beta,\delta,\gamma,R)$. All of these equations give the same answer for $\Phi_\rom{M}$. Moreover, one then checks that the Smarr relation holds. This procedure of using the first law to derive a physical quantity has also been used previously, e.g., in \cite{Gibbons:2004ai, Chong:2005hr}.

Now that we presented all conserved quantities, let us discuss the role of the parameter $\gamma$ in our general solution. This parameter was introduced as a boost performed after generating the electric and magnetic charges; see the form of the generator \eqref{generatork}. We found that the final solution has a momentum $P_z$ which does not vanish when $\gamma = 0$. This means that the solution has  already been boosted along the $z$-direction as a result of the intricate action of the $\mathfrak g_{2(2)}$ generators\footnote{We cannot resist speculating that this might be a supergravity reflection of the fact that when M2 charges are present on top of the M5 charges, the oscillator levels of the MSW CFT are shifted by an amount proportional to the M2 charges.}. It is however straightforward to find the boost $\gamma = \gamma_c(\beta,\delta)$ one has to apply to reach the non-boosted frame. Solving for $P_z = 0$, one finds
\be
\gamma_c(\beta,\delta) = \frac{1}{2} \tanh^{-1}\left[ \frac{2 \tanh 2 \beta}{(1+ 3 s_\delta^2)\sqrt{1 + 3 c_\delta^2}}\right] .\label{amitabhsigma}
\ee
In order to keep the boost as a free parameter, we then simply define the new boost parameter $\sigma$ as
\be
\sigma := \gamma_c(\beta,\delta) - \gamma.
\ee
The momentum  $P_z$ then takes the form
\begin{equation}
P_z = m c_{\sigma }s_{\sigma } \left(\frac{  \left(3 s_{\delta }^2+1\right)
   \left(c_\beta^2
   \sqrt{\frac{3 s_{\delta }^2}{2}+4} \left(3 s_{\delta
   }^2+2\right)-\frac{32 c_{\beta }^2 s_{\beta
   }^2}{\sqrt{3 c_{\delta }^2+1} \left(2 s_{\beta
   }^2+1\right) \left(3 s_{\delta
   }^2+1\right)}\right)}
{4 G_4 \sqrt{27 c_{\delta }^4
   s_{\delta }^2+\frac{4}{c_\beta^4}}} \right),
\end{equation}
i.e., $P_z$ is proportional to $c_\sigma s_\sigma$, where $c_\sigma = \cosh \sigma,$ and $s_\sigma = \sinh \sigma$.

\setcounter{equation}{0}
\section{Limits of the general string}
\label{sec:limits}

The general string \eqref{monster} is both electrically and magnetically charged and has finite temperature. In what follows we discuss simpler solutions that can be obtained either in the zero-temperature limits or in the limits when either the electric or the magnetic charge vanishes.

\subsection{Extremal limits}
\label{extremal_limits}
Any regular extremal limit with finite entropy per unit length  has $F[c_\gamma,s_\gamma]$ finite in \eqref{eq:A}. Since the temperature has the form \eqref{eq:T}, extremality is therefore equivalent to $r_+ = r_-$, which is achieved by the maximal rotation in the transverse space, equivalent to $a = \pm m$. Important exceptions however arises when $a$, $m$ both go to zero at different rates, while still implying $r_+ = r_-$. After a careful analysis, we distinguish three extremal limits
\begin{itemize}
\item[$(i)$] $a = \pm m ,\qquad m > 0$,
\item[$(ii)$] $ a = 0, \quad m\rightarrow 0, \quad \beta \rightarrow \infty, \quad$ such that
\be
m e^{2 \beta} = 2 P,
\ee
with $P$ finite and  keeping $\gamma$ and $\delta$ arbitrary free parameters.
\item[$(iii)$] $ a = 0, \quad m\rightarrow 0, \quad  \beta \to \infty ,\quad \delta \to 0, \quad  \gamma \to \infty,  \quad$  such that
\be
m e^{2 \beta} = 2q, \qquad  m e^{\gamma} = \frac{4 \sqrt{-\Diamond}}{q}, \qquad \frac{\delta}{m} = \frac{Q}{3\sqrt{-\Diamond}},
\ee
with $q$, $Q$ and $\Diamond$ finite, and $\Diamond < 0$.
\end{itemize}
In case $(i)$, the four-dimensional extremal solution is maximally rotating in the transverse space. Its explicit metric, gauge field and thermodynamics are trivially obtained from the previous section. Case $(ii)$ is nothing else than the three-parameter supersymmetric $\rom{M2^3-M5^3-P}$ string described in \cite{Kim:2010bf}\footnote{The fact that the parameters $\gamma$ and $\delta$ of the general string should be left free in the supersymmetric limit can be expected from the analysis of \cite{Elvang:2004xi, Emparan:2008qn}, where in the context of five-dimensional U(1)$^3$ supergravity a similar thing happens. We thank Roberto Emparan for clarifying this point.} (see also \cite{Bena:2005ni} or in the $N =2$, $d=4$ language \cite{Shmakova:1996nz})  that we describe in more details below\footnote{We thank Jan Perz for bringing reference \cite{Shmakova:1996nz} to our attention, and an anonymous  referee for drawing our attention to the results of section 4 of reference \cite{Bena:2005ni}.}. Finally, case $(iii)$ corresponds to the non-supersymmetric extremal string of \cite{Kim:2010bf} with independent $\rom{M2^3-M5^3-P}$ charges.

\subsubsection{Supersymmetric limit}
\label{app:extremal1}

The string in limit $(ii)$ is the $\rom{M2^3-M5^3-P}$ supersymmetric string of \cite{Kim:2010bf} and is connected to the supersymmetric $\rom{M5}^3-\rom{P}$ string. In $N=2,~D=4$ language the corresponding black hole solution was previously known in the literature \cite{Shmakova:1996nz}. (See also \cite{Bena:2005ni}.) For the benefit of the reader we review salient results from \cite{Bena:2005ni, Kim:2010bf}.

To set notation we very briefly review the general results first. Supersymmetric solutions of minimal supergravity      that we are interested in can be written as \cite{Gauntlett:2002nw}
\begin{equation}
\label{eqn:metric5dto4dEuclidean}
ds_5^2 = -f^2 (dt + \omega)^2+f^{-1} ds_4^2(\mathcal{B}),
\end{equation}
where $f$ and $\omega$ are a function and a one-form on the Gibbons-Hawking base space
with metric
\begin{equation}
\label{eqn:GibbonsHawkingBaseSpace}
ds_4^2(\mathcal{B}) = H^{-1}\left( d z + \chi \right)^2 + H ds_3^2.
\end{equation}
where $\star_3 d H =  d \chi.$ Defining  $G^+$ and $G^-$ as the self-dual and anti-self-dual parts of the form $f d\omega$ with respect to the Gibbons-Hawking metric $f d\omega = G^+ + G^-$, the field strength for supersymmetric spacetimes can be written as 
\begin{equation}
\label{eqn:FieldStrength}
F = -\sqrt{3} \,d\big(f (dt+\omega)\big)+\frac{2}{\sqrt{3}} G^+.
\end{equation}
Furthermore, one can write
\begin{equation}
\label{eqn:generalomega}
\omega = \omega_5 (d z+ \chi) + \omega_i dx^i ,
\end{equation}
where $\omega_5$ and $\omega_i$ are functions on $\mathbb{R}^3$. $\omega_i$ can be shown to be  directly related to the Lorentzian Taub-NUT charge for spacetimes that describe four-dimensional asymptotically flat black holes after reduction over the Gibbons-Hawking fibre $z$.

For the solution we report $\omega_i$ is identically zero because its has no Lorentzian NUT charge. Also, since we describe a five-dimensional black string, the harmonic $H$ is a constant. Hence, we fix $\chi = 0$. The functions $\omega_5$ and $f$ can be written in terms of $H$ and three additional harmonic functions as 
\begin{equation}
\label{eqn:fharmonic}
f^{-1} = K^2 H^{-1}+L, \qquad \omega_5 = H^{-2}K^3+\frac{3}{2} H^{-1}K L + M.
\end{equation}
For our solution the harmonic functions can be chosen to be
\bea
H &=&\frac{Q}{S},\qquad K =  \frac{\Delta}{S}+ \frac{q}{2r},\label{harm1} \\
L  &= &\frac{Q}{S}+ \frac{Q}{r}, \qquad M =  -\frac{1}{2}\left(\frac{\Delta}{S}+ \frac{3}{2r}(2\Delta -q)\right),\label{harm2}
\eea
where $S = \sqrt{Q^2+\Delta ^2}$. Another useful form of this solution can be found in \cite{Kim:2010bf}. 

In this parameterization, the three independent parameters take the form
\bea
q &=& \frac{4 P c_\delta^2 }{\sqrt{1 + 3 c_\delta^2}},\\
Q&=& 2 P c_\delta s_\delta \left( c_\gamma + \frac{s_\gamma}{\sqrt{1 + 3 c_\delta^2}}\right),\\
\Delta &=& \frac{2P}{3} \left( \frac{c_\gamma^2 + s_\gamma^2 + 3 c_\delta^2}{\sqrt{1+ 3 c_\delta^2}} - c_\gamma s_\gamma (1+ 3 s_\delta^2)\right).
\eea
The non-zero charges are
\begin{align}
Q_2 &= Q, & P_2 &= \frac{q}{2}, & Q_1 &= 3 \Delta -\frac{3 q}{2},\nn\\  G_4 M
&= \frac{3}{4} S,&
\Sigma &= -\frac{S^2 + 2 \Delta (\Delta -q)}{2 S}, & \Xi &= \frac{Q (\Delta -q)}{S}.
\end{align}
Note that $Q_2$, $P_2$ and $Q_1$ can be identified with the numerators of $1/r$ terms (up to numerical factors) in the harmonic functions \eqref{harm1}-\eqref{harm2}. The quartic invariant for the corresponding four-dimensional black hole is
\be
\Diamond = \frac{3}{4} q^2 \left(Q^2 -q^2+2 \Delta  q\right),
\ee
where the parameters $q, \Delta, Q$ are restricted to ensure the positivity of the quartic invariant. The ADM tension for the solution is
\be
\cT = \frac{1}{8 \pi R G_4 }\left(3 \Sigma_s  + 2G_4 M \right) = \frac{3 }{8 \pi R G_4}\frac{(q-\Delta ) \Delta}{ \sqrt{Q^2+\Delta ^2}}.
\label{tension}
\ee
When $\Delta =0$ or $\Delta = q$ the string becomes pressureless. In \cite{Kim:2010bf} and above the metric is presented in the Gibbons-Hawking form in order to facilitate comparison with \cite{Gauntlett:2002nw, Bena:2004wv, Elvang:2004rt, Bena:2005ni}. As a result the matrix $\cM$ does not go to identity at spatial infinity for the form of the metric given above, see section 3.4 of \cite{Kim:2010bf} for details.  In order to make sure that the matrix $\cM$ goes to identity at spatial infinity and to compare with the extremal limit of our general string in limit $(ii)$, the following shift of the coordinates is required in \eqref{eqn:metric5dto4dEuclidean}--\eqref{eqn:FieldStrength}
\be
z \rightarrow  z + \frac{\Delta }{\sqrt{Q^2+\Delta ^2}} \, t .
\ee
Finally note that the solution expressed in the coordinates $t^\prime =\lambda t$, $z^\prime = z/\lambda$ still has the form \eqref{eqn:metric5dto4dEuclidean}-\eqref{eqn:generalomega} with the harmonic functions \eqref{harm1}-\eqref{harm2} transformed as
\be
H^\prime = H/\lambda,\qquad L^\prime = \lambda L,\qquad K^\prime = K,\qquad M^\prime =\lambda^2 M .
\ee
One can then more easily take the $Q \rightarrow 0$ limit after having performed the above change of coordinates with $\lambda = Q/S$.

\subsubsection{Non-supersymmetric non-rotating limit}
\label{app:extremal2}
The three parameter extremal non-supersymmetric
black" into "extremal non-supersymmetric non-rotating black string of minimal supergravity with independent $\rom{M2^3-M5^3-P}$ charges \cite{Kim:2010bf} is most conveniently presented as
\bea
ds^2 &=& V^2 ds^2_3 + 2 V^{-1} dt dz + V^{-4} g dz^2, \label{nonsusymetric3} \\
A &=& - \frac{\sqrt{3}}{2} q \cos \theta d \phi - \frac{ \sqrt{3} Q}{r}\left(1+ \frac{q}{4r} \right) V^{-2}   dz, \label{nonsusymaxwell3}
\eea
where
\bea
V &=& 1 + \frac{q}{2r}, \\
g &=& 1+\frac{3 \Delta }{r}-\frac{3\left(q^2-3 \Delta  q+Q^2\right) }{2 r^2} + \frac{9 q^2 \Delta -4 q \left(q^2+Q^2\right) }{4r^3} -\frac{\Diamond}{4r^4}.
\eea
The quartic invariant for the corresponding four-dimensional black hole is
\be
\Diamond = - \frac{3}{4} q^2 \left(2 \Delta  q - q^2 - Q^2\right), \label{quarticnonsusy}
\ee
and the parameters $q, \Delta, Q$ are restricted to ensure that the quartic invariant is strictly \emph{negative}.
The non-zero charges in this parameterization are
\be
G_4 M = \frac{3 \Delta }{4}, \quad Q_1= \frac{3}{2} (q-2 \Delta ), \quad Q_2 = Q, \quad P_2  = \frac{q}{2}, \quad \Sigma = q-\frac{3 \Delta }{2}, \quad \Xi = -Q.
\ee
The ADM tension for the solution is
\be
\cT = \frac{3}{8\pi R G_4} (q - \Delta).
\ee
When $\Delta = q$ the string becomes pressureless. Again, the metric and gauge field as presented in \eqref{nonsusymetric3}--\eqref{nonsusymaxwell3} are such that the matrix $\cM$ does not go to identity at spatial infinity. In order to make sure that the matrix $\cM$ goes to identity at spatial infinity the shift of the $z$ coordinate $z \rightarrow z - t$ is required in \eqref{nonsusymetric3}--\eqref{nonsusymaxwell3}. One can then check that the resulting solution is identical to limit $(iii)$ of the general string \eqref{monster}--\eqref{monster_metric}.

\subsection{Spinning $\rom{M2}^3$ string}
\label{electric}

The boosted spinning $\rom{M2}^3$ string is obtained by setting $\beta = 0$ which cancels the magnetic charge $Q_\rom{M}$. The spacetime fields are presented in \cite{Compere:2009zh}, where  a detailed discussion of physical properties of the solution is also included, so we omit the details here. After setting $\beta = 0$, $\gamma$ can directly be regarded as the boost parameter along the string. When the boost parameter $\gamma$ is non-zero the extremal limit of the string is never supersymmetric.  When $\gamma$ is zero the solution admits a limit to the supersymmetric $\rom{M2}^3$ string. It is obtained by taking $a=0$, $m \rightarrow 0, \delta \rightarrow \infty$ such that $m e^{2 \delta} = 2 Q. $ In this limit, the conserved charges are
\be
G_4 M = \frac{3Q}{4}, \qquad Q_2 = Q, \qquad \Sigma = -\frac{Q}{2}
\ee
and the charge matrix becomes nilpotent and belongs to the $\mathcal{O}_2$ nilpotent $\tilde K$-orbit in the nomenclature of \cite{Kim:2010bf}.

\subsection{Spinning $\rom{M5}^3$ string}
\label{magnetic}

The boosted spinning $\rom{M5}^3$ string is obtained by setting $\delta = 0$ which cancels the electric charge $Q_\rom{E}$. The spacetime fields are again presented in \cite{Compere:2009zh}, where  a detailed discussion of physical properties of the solution is also included, so we omit the details here.
After setting $\delta = 0$,
\be
\sigma:= \beta - \gamma
\ee
can simply be regarded as the boost parameter along the string. The string admits a supersymmetric limit even when the boost parameter $\sigma$ is non-zero.  It is obtained by taking $a = 0$, $ m \rightarrow 0$, $\beta \rightarrow \infty,$ $\sigma \rightarrow \infty$ such that $m e^{2 \beta} = 2 P,$ and  $m e^{2 \sigma} = 2 Q.$ The conserved charges in this limit are
\be
G_4 M = \frac{|Q| + 3|P|}{4}, \qquad Q_1 = Q, \qquad P_2 = P, \qquad \Sigma = \frac{|P|-|Q|}{2}
\ee
and the corresponding charge matrix is nilpotent. The quartic invariant of the corresponding four-dimensional black hole is
\be
\Diamond = 4 QP^3.
\ee
For $P, Q> 0$ the quartic invariant is strictly positive, and the charge matrix belongs to the $\mathcal{O}_{3K}$ orbit in the nomenclature of \cite{Kim:2010bf}. The corresponding string is supersymmetric.  This string solution for $Q = 3 P$ corresponds to the infinite radius limit of the extremal dipole black ring of \cite{Emparan:2004wy}. For $P > 0$, $Q < 0$, (or $ P < 0$, $Q > 0$) the quartic invariant is strictly negative, and the charge matrix belongs to the $\mathcal{O}_{4K}'$ orbit in the nomenclature of \cite{Kim:2010bf}.   The corresponding string completely breaks supersymmetry. This solution for $Q = - 3 P$ also corresponds to the infinite radius limit of the extremal dipole black ring of \cite{Emparan:2004wy}, but with opposite sense of rotation in the plane of the ring. When $Q= 0,$ $P \neq 0$ the charge matrix belongs to the $\mathcal{O}_2$ orbit, and finally, when $Q \neq 0,$ $P = 0$ the charge matrix belongs to the $\mathcal{O}_1$ orbit \cite{Kim:2010bf}.

\setcounter{equation}{0}
\section{Decoupling limit and comparison with Larsen \cite{Larsen:2005qr}}
\label{sec:decoupling}

In this section we match the entropy of our general black string \eqref{monster}--\eqref{monster_metric} in the near-extremal limit with the prediction of \cite{Larsen:2005qr} in the context of the MSW CFT \cite{Maldacena:1997de}. In section \ref{decoupling} we discuss how to take the decoupling limit and in section \ref{matching} we match the entropy with the CFT prediction.

\subsection{Decoupling limit}
\label{decoupling}
Following \cite{Maldacena:1997re, Cvetic:1999ja}, the decoupling limit of our black string  \eqref{monster}--\eqref{monster_metric} is obtained by taking the five-dimensional Planck length to zero, $\ell_p \rightarrow 0$, while keeping the following quantities fixed:
\begin{itemize}
\item[$(i)$] We substitute the coordinate $r$ and the variables $m$ and $a$ as follows
\beq
r \rightarrow \tilde r {\ell_p^3 \over R^2}, \hspace{1cm}m \rightarrow \tilde{m} \,{\ell_p^3 \over R^2}, \hspace{1cm} a \rightarrow \tilde{a} \, {\ell_p^3 \over R^2}
\eeq
and keep $\tilde r$, $\tilde a$ and $\tilde m$ finite as $\ell_p \rightarrow 0$.  The radius of the string direction $R$ is introduced in these rescalings to keep the quantities $\tilde r$, $\tilde m$, $\tilde a$ dimensionless.  We also introduce the rescaled coordinates $\tau  = \frac{n \ell_p t }{R}$ and $\sigma = \frac{z}{R}$.
\item[$(ii)$]  We keep the number of M5 and M2 branes, $n$ and $N$ respectively, fixed as $\ell_p \to 0$. In the near extremal limit we require that the number of M5 branes is much larger than the number of anti-M5  branes.  Assuming that the supergravity solution can be modeled by free M2 and M5 branes, the number of branes is then given by following quantization (see e.g. \cite{Emparan:2006mm})
\be
N = \frac{1}{\sqrt{3}} \left( \frac{4G_5}{\pi} \right)^{1/3} Q_\rom{E} = \frac{\ell_p Q_\rom{E}}{\sqrt{3}}, \quad  \quad n= \frac{2}{\sqrt{3}} \left( \frac{\pi}{4 G_5} \right)^{1/3} Q_\rom{M} = \frac{2Q_\rom{M}}{\sqrt{3} \ell_p}.
\ee
\item[$(iii)$] We  quantize linear momenta $P_z$ in integer units $n_p$ of $\frac{1}{R}$ and keep $n_p$ fixed as $\ell_p \rightarrow 0$,
\be
n_p = R P_z = \frac{2 \pi R^2 Q_1}{4 G_5} = \frac{2 R^2 Q_1}{\ell_p^3}, \hspace{1cm} n_p \in \mathbb{Z}.
\ee
\end{itemize}

This limit also corresponds to a near horizon limit since $r,r_{\pm} \rightarrow 0$ as $\ell_p \rightarrow 0$. In terms of the parameters $\beta, \delta, \gamma$ of our general string this limit is achieved as
\be
\ell_p^2 e^{2\beta} \rightarrow \frac{n R^2}{\tilde m}, \quad \ell_p^{-1}\delta e^{\gamma} \rightarrow \frac{2 N}{3 n R}, \quad \ell_p^{-2}e^{-2\gamma} \rightarrow  \frac{B}{2 n R^2},
\ee
where 
\beq
B=  n_p + \frac{3 N^2}{4n} + \sqrt{4 \, \tilde{m}^2 + \left( n_p + \frac{3 N^2}{4n}\right)^2}  \label{bigB}
\eeq
is non-negative. In this limit  we obtain a BTZ black hole times a two-sphere sphere from the general string \eqref{monster}--\eqref{monster_metric},
\beq
ds^2_{\mbox{\tiny decoupling}} =  ds^2_{\rom{BTZ}} + ds^2_{\rom{S}^2},
\eeq
with
\bea
ds^2_{\rom{BTZ}}  &=& - N^2 d \tau^2 + N^{-2} d\rho^2 + \rho^2 (d \sigma + N_\sigma d \tau)^2, \\
N^2 &=& \frac{\rho^2}{R^2_\rom{AdS}} - 8 G_3 M_3 +  \frac{16 G_3^2 J_3^2}{\rho^2}, \quad N_\sigma =  \frac{4G_3 J_3}{\rho^2}
\eea
and
\be
ds^2_{\rom{S}{}^2} = R_{\rom{S}{}^2}^2 ( d\theta^2 + \sin^2\theta d\tilde \phi)^2,
\ee
and where the new radial coordinate $\rho$ is related to $\tilde r$ as 
\be
\tilde r = \tilde m - \frac{n_p}{2}-\frac{3 N^2}{8n}+\frac{\tilde a^2 - 2 \tilde m^2}{B} + \frac{n}{2 l_p^2}\rho^2.
\ee
To get the decoupled metric in the above factorized form the following shift of the $\phi$ coordinate was required
\be
\tilde \phi = \phi -  \frac{2 \sqrt{2} \tilde a}{R \sqrt{n^3 B}} \left( t - z \right).
\ee
This shift corresponds to a spectral flow from the CFT point of view. In the decoupling limit, the gauge field admits a magnetic flux on the two-sphere 
\be
A = -\sqrt{3}R_{\rom{S}^2}\cos\theta d \tilde\phi.
\ee
In the decoupled geometry, the radii of the three-dimensional AdS space and the two-sphere are $ R_{\rom{AdS}} = 2 R_{\rom{S}^2} = n \ell_p,$ and the mass and the angular momentum parameters of the BTZ black hole are,
\beq
M_3 &=& \frac{1}{4 n^2 \ell_p}\left( 4 n n_p + 3 N^2 + \frac{ 8 n(2 \tilde{m}^2  - \tilde{a}^2)}{B} \right),  \\
J_3 &=&   n_p + {3 N^2 \over 4 n} + { 2 \tilde{a}^2\over B}.
\eeq
In arriving at these expressions we have related the effective three-dimensional Newton's constant  to the five-dimensional as
\be
G_3  = \frac{G_5}{A_2}  = \frac{\ell_p}{4 n^2},
\ee
where $A_2$ is the area of the two-sphere. The central charge of the CFT induced on the boundary of AdS$_3$ is given by
\beq
c= {3 R_\rom{AdS} \over 2 G_3 }= 6 n^3 , \label{cc}
\eeq
and the conformal weights $h_L^{\rom{irr}}$ and $h_R^{\rom{irr}}$ (eigenvalues of the Virasoro operators $L_0$ and $\bar L_0$, respectively) are related to the BTZ parameters as,
\bea
h_\rom{L}^{\rom{irr}} &=& \frac{(R_\rom{AdS} M_3 +  J_3)}{2} = \frac{B}{2}, \nn \\
h_\rom{R}^{\rom{irr}} &=& \frac{(R_\rom{AdS} M_3 -  J_3)}{2} = \frac{2 (\tilde m^2-\tilde a^2)}{B}. \label{conformalweights}
\eea
In the next section we will see that the conformal weights \eqref{conformalweights} exactly match with a statistical-mechanical counting based on the MSW CFT.

We can now deduce the possible extremal limits of our black string \eqref{monster}--\eqref{monster_metric} by setting either $h_\rom{L}^{\rom{irr}}$ and $h_\rom{R}^{\rom{irr}}$ to zero.
Let us first consider $h_\rom{R}^{\rom{irr}} =0$. It vanishes only if $\tilde a = \pm \tilde m $, and $B \neq 0$, which corresponds to the general extremal limit---case $(i)$ of section \ref{extremal_limits}. Moreover, if we set $\tilde a = 0$ and require $(4 n n_p + 3 N^2)$ to be positive and take $\tilde m \to 0$, we recover the extremal supersymmetric limit---case $(ii)$ of section \ref{extremal_limits}. Finally,
let us consider $h_\rom{L}^{\rom{irr}} =0$. For it to vanish we need $B$ to vanish while keeping $h_\rom{R}^{\rom{irr}}$ finite. This can only be achieved by setting $\tilde a= 0$ with $(4 n n_p + 3 N^2)$ negative as we take $\tilde m \to 0$. In this limit we recover the non-supersymmetric extremal string---case $(iii)$ of  section \ref{extremal_limits}.

\subsection{Matching with the CFT prediction}
\label{matching}

For extremal black rings the two-dimensional $(4_R,0_L)$ Maldacena-Strominger-Witten CFT \cite{Maldacena:1997de} has been proposed as the low energy theory for the microscopics \cite{Emparan:2004wy,Cyrier:2004hj,Bena:2004tk}.  This identification is motivated by the fact that the $S^1\times S^2$ topology of the horizon of the ring is same as that of a black string,
which is well described by the MSW CFT.  In  \cite{Larsen:2005qr}, working under the hypothesis that near-extremal black rings can also be identified
with black strings, an entropy formula of the near-extremal black ring was proposed. Our general black string is precisely the supergravity black string on which these considerations are based, though at the time of \cite{Larsen:2005qr} this solution was not explicitly known.
It is thus expected that the CFT result should reproduce the Bekenstein-Hawking entropy of our string in the near-extremal limit. In this section we show that this is indeed the case.

The CFT expression for entropy is given  as usual by the Cardy formula
\be
S = 2 \pi \left( \sqrt{ c \, h_{\rom{L}}^{\rom{irr}}\over 6}+\sqrt{ c \, h_{\rom{R}}^{\rom{irr}}\over 6}\right). \label{entropylarsen}
\ee
where the central charge $c$ is
\be
c= 6 n^3, \label{central}
\ee
$n$ being the (equal) number of M5 branes along the cycles orthogonal to the canonical (12) (34) and (56) two cycles. See brane intersection Table \ref{braneintersection}. This value of the central charge matches \eqref{cc}. In \eqref{entropylarsen} $h_{\rom L}^{\rom{irr}}$ and $h_{\rom R}^{\rom{irr}}$ are respectively the left and right moving oscillator levels of the CFT. In the MSW CFT, M2 branes are realized as charged excitations, and as a result  the oscillator levels are shifted by an amount proportional to the M2 charges. A clear derivation of these results is found in \cite{Larsen:2005qr}; here we simply quote the results from there:\beq
h_{\rom L}^{\mbox{\tiny irr}} &=& {\epsilon + n_p \over 2},\\
h_R^{\mbox{\tiny irr}} &=& {\epsilon - n_p \over 2}- {3 N^2\over 4 n} -  \frac{J^2}{4 n^3},
\eeq
where $N$ is the (equal) number of M2 wrapping (12) (34) and (56) two cycles, $\epsilon$ is the excitation energy, $n_p$ is the momentum quantum number along the $z$-direction, and, finally, $J$ is the quantum number under a U(1) subgroup of the SU(2) R-symmetry group. The SU(2) R-symmetry is interpreted as the rotation group in the four-dimensional spacetime transverse to the string; therefore, $\frac{J}{2}$ is simply the angular momentum in the four-dimensional spacetime transverse to the string. Note that with respect to reference \cite{Larsen:2005qr}, we have set $q^1=q^2=q^3=n$ and $Q_1=Q_2=Q_3=N$, because we are considering minimal five-dimensional supergravity. 
\begin{table}
\bea
\begin{array}{|c| c| c c c c c c |c| c c c|}
\hline
     & t   &   z_1  &    z_2&   z_3  &   z_4  &   z_5   & z_6    &     z &    r   &   \theta  & \phi  \\
     \hline 
  \rom{M2} & -   &    -   &      -& \sim  &  \sim & \sim   & \sim  &  \sim &  \cdot &   \cdot   & \cdot \\
  \rom{M2}  & -   & \sim  & \sim &   -    &   -    & \sim   & \sim  &  \sim &  \cdot &   \cdot   & \cdot \\
  \rom{M2}  & -   & \sim  & \sim & \sim  &  \sim & -       & -      &  \sim &  \cdot &   \cdot   & \cdot \\
  \hline 
  \rom{M5}   & -   & \sim  & \sim & -      &  -     &   -     &   -    &  -    &  \cdot &   \cdot   & \cdot \\
  \rom{M5}  & -   &    -   &      -& \sim  &  \sim &  -      &    -   &  -    &  \cdot &   \cdot   & \cdot \\
  \rom{M5}  & -   &    -   &      -&  -     &   -    & \sim   & \sim  &   -   &  \cdot &   \cdot   & \cdot \\
  \hline
  \rom{P} & -   & \sim  & \sim & \sim  &  \sim & \sim   & \sim &- &  \cdot &   \cdot   & \cdot \\
  \hline
\end{array}
\label{M2M5}
\eea
\caption{Following \cite{Peet:2000hn}, a $-$ indicates a worldvolume direction of the brane, a $\sim$ indicates that the brane is smeared in that direction, while a dot $\cdot$ indicates that the brane is pointlike in that direction.\label{braneintersection}}
\end{table}

Now we would like to express the CFT quantities in term of our supergravity quantities. The variable $\epsilon$ should be interpreted as the energy above the BPS ground state quantized in units of $\frac{1}{R}$. From the quantization condition of the charge $P_2$ we have
\beq n= 2 \left({\pi \over 4 G_5}\right)^{1/3} P_2 =  \frac{2}{\ell_p} P_2.
\eeq
This implies that, in our conventions, a single M5 brane carries the charge
$P_2^\rom{unit}=\frac{\ell_p}{2}$. 
The mass of the single M5 brane is $M^\rom{unit}_{\rom{M5}} = {R \over \ell_p^2}.$ Therefore, we conclude that $\epsilon$ is
\beq
\epsilon = R (M_{5} - 3 M_{\rom{ M5}})
\hspace{1cm}
\mbox{where}
\hspace{1cm}
M_{\rom{M5}}= n  M^\rom{unit}_{\rom{M5}}  = {n R \over \ell_p^2}.
\eeq
Explicitly, we have
\beq
\epsilon =  \frac{3 N^2}{4n} + \sqrt{4 \tilde m^2  + \left(n_p + \frac{3 N^2}{4n}\right)^2} = B - n_p,
\eeq
in the strict decoupling limit, where $B$ is given by \eqref{bigB}. 
The half-integer quantized angular momentum $J_\phi$ is half the integer quantized charge $J$. We find,
\be
\frac{J^2}{4 n^3} = \frac{J_{\phi}^2}{n^3} = \frac{2 \tilde a^2}{B}.
\ee
Putting this all together we immediately see that the CFT result precisely matches with the conformal weights $h_L^{\rom{irr}}$ and $h_R^{\rom{irr}}$ \eqref{conformalweights} of the previous section
\bea
h_\rom{L}^{\rom{irr}} &=&  \frac{B}{2}, \nn \\
h_\rom{R}^{\rom{irr}} &=& \frac{2 (\tilde m^2-\tilde a^2)}{B}. \label{conformalweights2}
\eea
Thus, the Bekenstein-Hawking entropy of our general string in the near-extremal limit is reproduced by the MSW CFT.

\setcounter{equation}{0}
\section{Discussion and implications for black rings}
\label{sec:discussion}

 In this paper we have presented the most general black string solution to ungauged minimal supergravity in five dimensions. Our solution has five independent parameters: energy above the BPS bound, angular momentum in the four-dimensional transverse space, linear momentum along the string, smeared electric zero-brane charge, and magnetic one-brane charge.  Our solution admits three separate extremal limits, one of which is supersymmetric and the other two are non-supersymmetric. We also discussed physical properties and thermodynamics of our string.  In the near-extremal limit, the entropy of our general black string exactly matches the CFT prediction.

The general black string solution presented in this paper has a number of implications for yet-to-be-found black rings in ungauged as well as gauged minimal five-dimensional supergravity. We discuss some of the important points in this section. A more complete analysis is left for the future.

\begin{figure}[htp]
  \begin{center}
\includegraphics[scale=0.75]{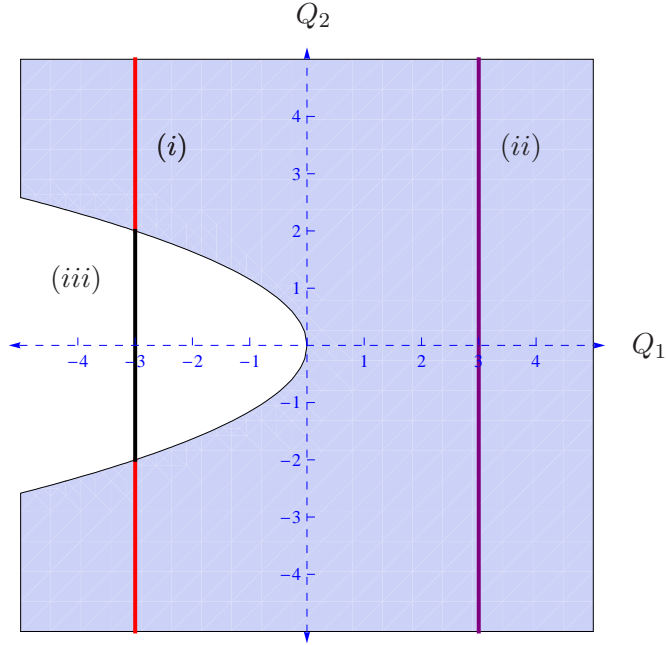}
\put(-170,185){$(i)$}
\put(-210,135){$(iii)$}
\put(-40,185){$(ii)$}
\put(-170,185){$(i)$}
\put(10,110){$Q_1$}
\put(-117,235){$Q_2$}
\caption{A working phase diagram for a class of extremal black rings of minimal ungauged supergravity. This diagram is obtained from extremal black strings within the assumptions of the blackfold approach; i.e., it is obtained by imposing pressureless condition on the extremal black strings of \cite{Kim:2010bf}. Internal rotation on strings is not considered in this diagram, although the corresponding black ring might have both angular momenta non-zero. The vertical axis is the M2$^3$ charge $(Q_2)$ and the horizontal axis is the boost charge $(Q_1)$, and the M5$^3$ charge ($P_2$) is held fixed in this diagram. The shaded region in this diagram (positive quartic invariant for $P_2 = 1$: $\Diamond = 3 Q_2^2 + 4 Q_1 >0$)   shows the parameter space for supersymmetric black string of \cite{Kim:2010bf}, and the unshaded region shows the parameter space for the non-supersymmetric black string of \cite{Kim:2010bf}. The pressureless condition is satisfied only on the two vertical lines; not anywhere else. Thus, the vertical lines show the allowed parameter space for extremal black rings for a fixed value of the dipole charge. Branch $(i)$ corresponds to the supersymmetric black ring of \cite{Elvang:2004rt}.  General solutions corresponding to branches $(ii)$ and $(iii)$ are yet to be discovered, though, in the limit when the M2$^3$ charge goes to zero, these solutions are known. They correspond to the clockwise and counterclockwise rotating extremal dipole black rings of \cite{Emparan:2004wy}. Note that the supersymmetric branch $(i)$ connects to the non-supersymmetric branch $(iii)$ as the M2$^3$ charge is reduced below the allowed range.}
\label{fig:phase}
\end{center}
\end{figure}

\paragraph{New extremal branches of black rings:}
In section \ref{app:extremal1} we noted that there are two pressureless limits of the three parameter supersymmetric black string of \cite{Kim:2010bf}; and in section \ref{app:extremal2} we noted that there is one pressureless limit of the three parameter extremal non-supersymmetric black string of \cite{Kim:2010bf}. The blackfold approach \cite{Emparan:2009cs, Emparan:2009at, Emparan:2009vd} suggests that all pressureless black strings describe the infinite radius limit of some black ring. Since the black strings of \cite{Kim:2010bf} are all extremal, the corresponding black rings will also be extremal. Furthermore, since considerations of \cite{Kim:2010bf} exhaust all possible extremal black strings of minimal ungauged supergravity with no rotation in the transverse space, the pressureless black strings of \cite{Kim:2010bf} provide a working phase diagram for a class of extremal black rings of this theory.
To draw this phase diagram let us collect important properties of extremal pressureless black strings from sections \ref{app:extremal1} and \ref{app:extremal2}. This is summarized in the table below. For concreteness we only consider $q > 0$; $q < 0$ can be dealt similarly.
\vskip 0.3cm
\begin{table}
\begin{center}
\begin{tabular}{|c|c|c|}\hline
$(i)$:  $\Delta = 0$ susy & $(ii)$:  $\Delta = q$ susy & $(iii)$:  $\Delta = q$ non-susy \\ \hline
$Q_2 = Q$ & $Q_2 = Q$ & $Q_2 = Q$ \\ \hline
$P_2 = \frac{q}{2}$ & $P_2 = \frac{q}{2}$ & $P_2 = \frac{q}{2}$ \\ \hline
$Q_1 = -\frac{3q}{2}$ & $Q_1 =  + \frac{3q}{2}$ & $Q_1 = -\frac{3q}{2}$ \\ \hline
$|Q_2| > q $ & no condition & $|Q_2|< q$ \\ \hline
$M_5 = \sqrt{3} |Q_E|$ & $M_5 = \sqrt{3} |Q_E| \sqrt{1 + \frac{q^2}{Q^2}}$ & $M_5 = \frac{3 \pi R q}{2G_5}$  \\ \hline
\end{tabular}
\caption{Properties of extremal pressureless black strings
\label{pressureless}}
\end{center}
\end{table}
\vskip 0.2cm
From Table \ref{pressureless} we can immediately draw a number of conclusions.
Clearly, only case $(i)$ saturates the five-dimensional BPS bound. It corresponds to the black string of \cite{Bena:2004wv}, which in turn corresponds to the infinite radius limit of the supersymmetric black ring of \cite{Elvang:2004rt}. In case $(i)$ the M2$^3$ electric charge $Q$ is bounded from below $|Q| > q$. As this bound is violated, solution $(i)$ ceases to be smooth. Precisely when solution $(i)$ becomes singular, the pressureless non-supersymmetric solution, case $(iii)$, takes over.   See Figure \ref{fig:phase}.  Branch $(iii)$ is completely smooth. On this branch one can continuously take the electric charge to zero. Thus the corresponding black ring is connected to an extremal dipole ring of \cite{Emparan:2004wy}. The dipole ring is such that in the infinite radius limit, the boost along the string completely breaks supersymmetry. Black ring solutions on this branch are not known in full generality. Case $(ii)$, on the other hand, indicates that there is another branch of extremal black rings of  minimal supergravity.
This branch is also connected to an extremal dipole ring of \cite{Emparan:2004wy}. The dipole ring on this branch is such that in the infinite radius limit, the boost along the string \emph{preserves} supersymmetry. Black ring solutions on this branch are also not known in full generality.

\begin{figure}[htp]
  \begin{center}
    \subfigure{\label{fig:non-susy}\includegraphics[scale=0.25]{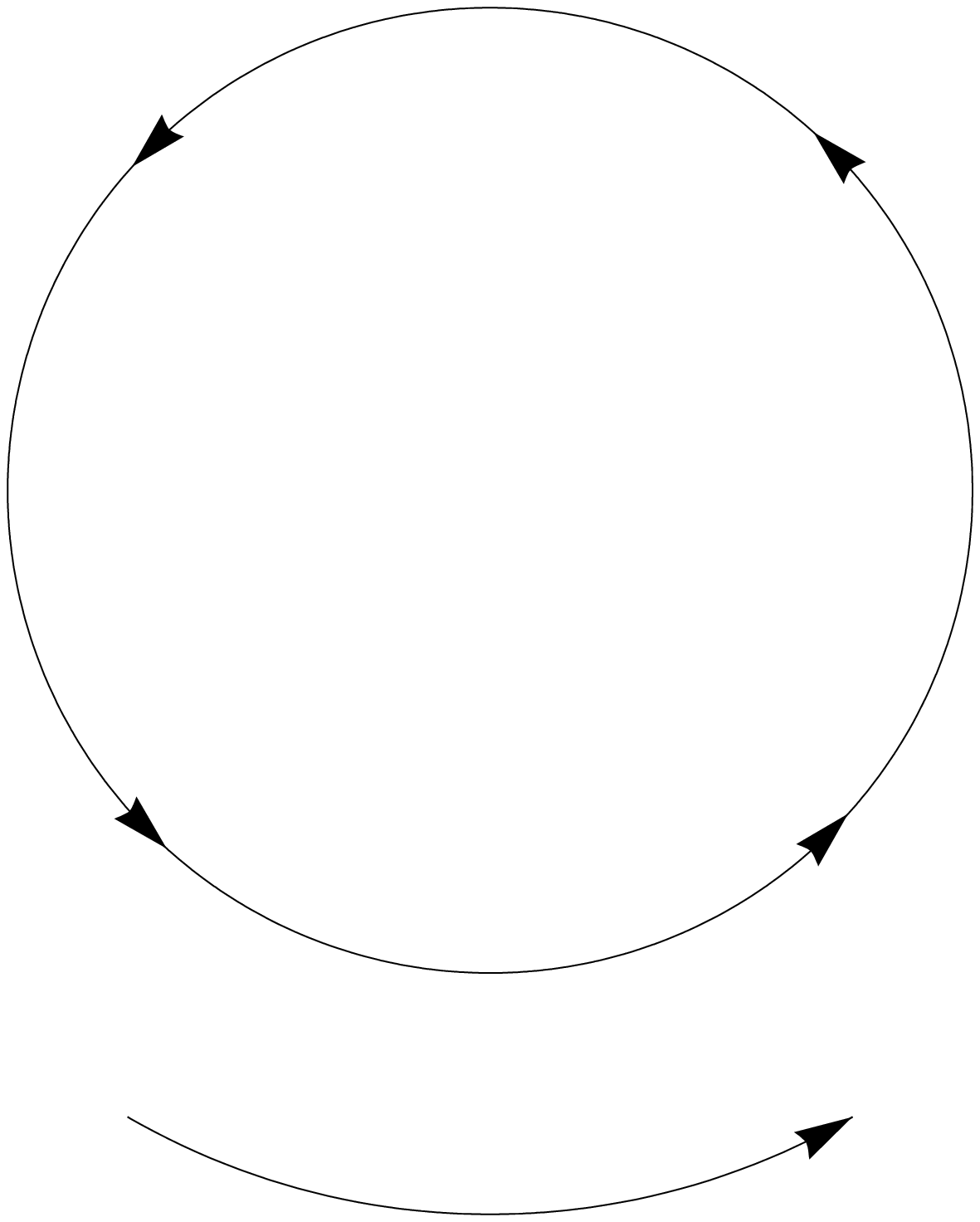}}
    \hskip 2.5cm
   \subfigure{\label{fig:susy}\includegraphics[scale=0.25]{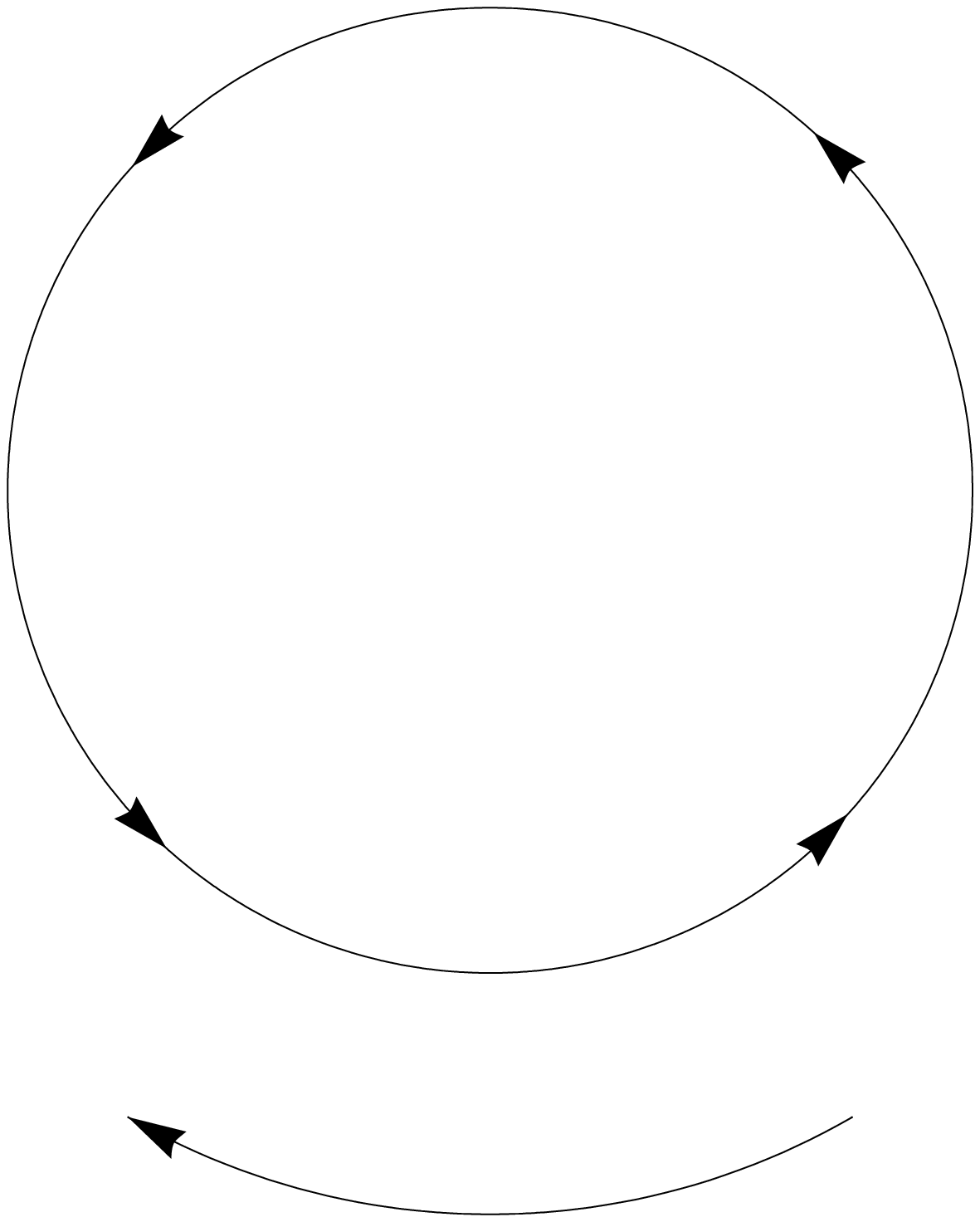}} \\
  \end{center}
  \caption{The two branches of pressureless strings bent into rings. The aligned string (left) has the same orientation for the momentum along the string and the magnetic charge, while the anti-aligned string (right) has the opposite relative orientation. In the extremal limit, the aligned string (branch $(ii)$ of figure \ref{fig:phase}) is connected to the infinite radius limit of a dipole ring. The dipole ring is such that it preserves supersymmetry  in the infinite radius limit. The anti-aligned string in the extremal limit is branch $(i)$ or branch $(iii)$ of figure \ref{fig:phase} depending on the value of the electric charge. Branch $(iii)$  is also connected to the the infinite radius limit of a dipole ring. The  dipole ring in this case breaks supersymmetry  in the infinite radius limit.}
  \label{fig:rings}
\end{figure}

\paragraph{Most general black ring:}

A five parameter family of black rings is conjectured to exist in minimal supergravity \cite{Elvang:2004xi}. For ranges of parameters for which this family would admit the infinite radius limit, the corresponding string solution would be contained in our general string. After some manipulations, one can show that for our general black string the tension \eqref{eq:T2} can be set to zero for two distinct values of the boost parameter when the remaining parameters are left arbitrary (and finite). The two distinct branches of pressureless strings are distinguished by the relative sign of $Q_1$ and $P_2$, or in five-dimensional language, by the property that the orientation of the momentum along the string is  aligned or anti-aligned with the orientation of the magnetic charge $Q_\rom{M}$. We refer to these two branches as the \emph{aligned} pressureless string and the \emph{anti-aligned} pressureless string. The former contains branch $(ii)$ of the pressureless extremal strings while the latter contains branches $(i)$ and $(iii)$ of the pressureless extremal strings. See Figure \ref{fig:rings}.

Let us examine the properties of these pressureless strings in a little more detail. Without loss of generality, we only consider $Q_\rom{E} \geq 0$ and $Q_\rom{M} \geq 0$ by choosing $\beta,\delta \geq 0$ as explained in section \ref{sec:monster}. The aligned string has then $P_z > 0$ and the anti-aligned string has $P_z <0$. After a careful numerical analysis, we find that the two branches obey the following bound
\be
\frac{\sqrt{2}}{3} M^{\rom{crit}}_5 \leq |P_z| \leq \frac{M_5}{2},
\ee
where $M^\rom{crit} = \frac{3\pi m R}{G_5}$ is the mass of the neutral pressureless black string ($\delta = \beta = 0$). The fact that the (absolute value of the) linear momentum is bounded from below indicates that the two branches are always separated. Also, we note that for a fixed value of the mass and magnetic charge, the two branches of pressureless strings have a different electric charge which always obeys
\be
Q_\rom{E}^\rom{aligned} \geq Q_\rom{E}^\rom{anti-aligned}, \qquad \text{at fixed } M_5, \; Q_\rom{M}, \;\text{and} \; \; |P_z|.
\ee
The equality holds only when there is no magnetic charge or when the electric charge simply vanish. It would be interesting to have an intuitive understanding of this physical effect.
The presence of two distinct pressureless limits suggests
that there are two
classes of black rings each admitting an infinite radius limit. Black rings on different branches rotate in different directions relative to the orientation of the dipole charge. See figure \ref{fig:rings}.

\paragraph{Extremal black rings in AdS:}

The analysis of  \cite{Kunduri:2006uh, Kunduri:2007qy} shows that there is no smooth near-horizon geometry that describes a supersymmetric black ring in AdS---the
near horizon geometry of a supersymmetric black ring in AdS necessarily possesses a conical singularity. In \cite{Caldarelli:2008pz} a more physical argument was presented for the absence of supersymmetric black rings in AdS. It was based on the observation that the AdS potential requires pressure along the one-brane in order to achieve mechanical equilibrium when the one-brane is curved into a contractible circle in AdS. It was speculated that supersymmetry is presumably incompatible with this pressure, and so, in AdS we should not expect supersymmetric black rings.
However, the three parameter supersymmetric black string of \cite{Kim:2010bf} invalidates the argument made in \cite{Caldarelli:2008pz}.  The supersymmetric black string of \cite{Kim:2010bf} can have non-zero pressure (see \eqref{tension}), so one can indeed balance it in AdS by curving it into a contractible circle. However, upon curving a straight supersymmetric string, the configuration will, in general, not remain supersymmetric.
The extremal strings of \cite{Kim:2010bf} and more generally our general string provide requisite candidate strings that might correspond to the infinite radius limit of some (non-)extremal black rings in AdS. A detailed perturbative study following \cite{Caldarelli:2008pz, Emparan:2007wm} is left for the future.

\subsection*{Acknowledgements}
We would like to thank St\'ephane Detournay, Roberto Emparan, Axel Kleinschmidt, Gary Horowitz and Jan Perz for discussions,  and Ella Jamsin for collaboration at initial stages of the project. This research is  supported in part by the National Science Foundation under Grant No. PHY05-51164.   AV is supported by IISN - Belgium (conventions 4.4511.06
and 4.4514.08) and by the Belgian Federal Science Policy Office through the Interuniversity Attraction Pole P6/11. The work of G.C. is supported in part by the US National Science Foundation under Grant No.~PHY05-55669, and by funds from the University of California. The work of S.dB. is funded by the European Commission though the grants PIOF-GA-2008-220338 (Home institution: Universit\'e Libre de Bruxelles, Service de Physique Th\'eorique et Math\'ematique, Campus de la Plaine, B-1050 Bruxelles, Belgium).
The work of SS is supported by the Natural Sciences and Engineering Research Council of Canada. SS would also like to thank the University of California, Santa Barbara for their gracious hospitality where this research was conducted.



\begin{thebibliography}{99}





\bibitem{Strominger:1996sh}
  A.~Strominger and C.~Vafa,
``Microscopic Origin of the Bekenstein-Hawking Entropy,''
  Phys.\ Lett.\  B {\bf 379}, 99 (1996)
  [arXiv:hep-th/9601029].


\bibitem{Mathur:2005ai}
  S.~D.~Mathur,
``The quantum structure of black holes,''
  Class.\ Quant.\ Grav.\  {\bf 23}, R115 (2006)
  [arXiv:hep-th/0510180].

\bibitem{David:2002wn}
  J.~R.~David, G.~Mandal and S.~R.~Wadia,
``Microscopic formulation of black holes in string theory,''
  Phys.\ Rept.\  {\bf 369}, 549 (2002)
  [arXiv:hep-th/0203048].


\bibitem{Peet:2000hn}
  A.~W.~Peet,
``TASI lectures on black holes in string theory,''
  arXiv:hep-th/0008241.



\bibitem{Callan:1996dv}
  C.~G.~Callan and J.~M.~Maldacena,
``D-brane Approach to Black Hole Quantum Mechanics,''
  Nucl.\ Phys.\  B {\bf 472}, 591 (1996)
  [arXiv:hep-th/9602043].


\bibitem{Horowitz:1996fn}
  G.~T.~Horowitz and A.~Strominger,
  ``Counting States of Near-Extremal Black Holes,''
  Phys.\ Rev.\ Lett.\  {\bf 77}, 2368 (1996)
  [arXiv:hep-th/9602051].



\bibitem{Breckenridge:1996sn}
  J.~C.~Breckenridge, D.~A.~Lowe, R.~C.~Myers, A.~W.~Peet, A.~Strominger and C.~Vafa,
  ``Macroscopic and Microscopic Entropy of Near-Extremal Spinning Black
  Holes,''
  Phys.\ Lett.\  B {\bf 381}, 423 (1996)
  [arXiv:hep-th/9603078].

\bibitem{Elvang:2004rt}
  H.~Elvang, R.~Emparan, D.~Mateos and H.~S.~Reall,
``A supersymmetric black ring,''
  Phys.\ Rev.\ Lett.\  {\bf 93}, 211302 (2004)
  [arXiv:hep-th/0407065].



\bibitem{Cyrier:2004hj}
  M.~Cyrier, M.~Guica, D.~Mateos and A.~Strominger,
``Microscopic entropy of the black ring,''
  Phys.\ Rev.\ Lett.\  {\bf 94}, 191601 (2005)
  [arXiv:hep-th/0411187].


\bibitem{Bena:2004tk}
  I.~Bena and P.~Kraus,
``Microscopic description of black rings in AdS/CFT,''
  JHEP {\bf 0412}, 070 (2004)
  [arXiv:hep-th/0408186].



\bibitem{Maldacena:1997de}
  J.~M.~Maldacena, A.~Strominger and E.~Witten,
``Black hole entropy in M-theory,''
  JHEP {\bf 9712}, 002 (1997)
  [arXiv:hep-th/9711053].




\bibitem{Emparan:2006mm}
  R.~Emparan and H.~S.~Reall,
``Black rings,''
  Class.\ Quant.\ Grav.\  {\bf 23}, R169 (2006)
  [arXiv:hep-th/0608012].


\bibitem{Emparan:2008qn}
  R.~Emparan,
``Exact Microscopic Entropy of Non-Supersymmetric Extremal Black Rings,''
  Class.\ Quant.\ Grav.\  {\bf 25}, 175005 (2008)
  [arXiv:0803.1801 [hep-th]].



\bibitem{Larsen:2005qr}
  F.~Larsen,
``Entropy of thermally excited black rings,''
  JHEP {\bf 0510}, 100 (2005)
  [arXiv:hep-th/0505152].




\bibitem{Cvetic:1996xz}
  M.~Cvetic and D.~Youm,
``General Rotating Five Dimensional Black Holes of Toroidally Compactified
Heterotic String,''
  Nucl.\ Phys.\  B {\bf 476}, 118 (1996)
  [arXiv:hep-th/9603100].




\bibitem{Elvang:2003mj}
  H.~Elvang and R.~Emparan,
``Black rings, supertubes, and a stringy resolution of black hole
non-uniqueness,''
  JHEP {\bf 0311}, 035 (2003)
  [arXiv:hep-th/0310008].


\bibitem{Elvang:2004xi}
  H.~Elvang, R.~Emparan and P.~Figueras,
``Non-supersymmetric black rings as thermally excited supertubes,''
  JHEP {\bf 0502}, 031 (2005)
  [arXiv:hep-th/0412130].


\bibitem{Compere:2009zh}
  G.~Compere, S.~de Buyl, E.~Jamsin and A.~Virmani,
``G2 Dualities in D=5 Supergravity and Black Strings,''
  Class.\ Quant.\ Grav.\  {\bf 26}, 125016 (2009)
  [arXiv:0903.1645 [hep-th]].


\bibitem{Bouchareb:2007ax}
  A.~Bouchareb, G.~Clement, C.~M.~Chen, D.~V.~Gal'tsov, N.~G.~Scherbluk and T.~Wolf,
``$G_2$ generating technique for minimal D=5 supergravity and black rings,''
  Phys.\ Rev.\  D {\bf 76}, 104032 (2007)
  [Erratum-ibid.\  D {\bf 78}, 029901 (2008)]
  [arXiv:0708.2361 [hep-th]].


\bibitem{Gal'tsov:2009da}
  D.~V.~Gal'tsov and N.~G.~Scherbluk,
 ``Three-charge doubly rotating black ring,''
  Phys.\ Rev.\  D {\bf 81}, 044028 (2010)
  [arXiv:0912.2771 [hep-th]].


\bibitem{Tanabe:2008vz}
  M.~Tanabe,
``The Kerr black hole and rotating black string by intersecting M-branes,''
  arXiv:0804.3831 [hep-th].



\bibitem{Kim:2010bf}
  S.~S.~Kim, J.~L.~Hornlund, J.~Palmkvist and A.~Virmani,
``Extremal solutions of the S3 model and nilpotent orbits of G2(2),''
  arXiv:1004.5242 [hep-th].




\bibitem{Breitenlohner:1987dg}
  P.~Breitenlohner, D.~Maison and G.~W.~Gibbons,
``Four-Dimensional Black Holes from Kaluza-Klein Theories,''
  Commun.\ Math.\ Phys.\  {\bf 120}, 295 (1988).




\bibitem{Duff:1995sm}
  M.~J.~Duff, J.~T.~Liu and J.~Rahmfeld,
``Four-Dimensional String-String-String Triality,''
  Nucl.\ Phys.\  B {\bf 459}, 125 (1996)
  [arXiv:hep-th/9508094].


\bibitem{Cremmer:1999du}
  E.~Cremmer, B.~Julia, H.~Lu and C.~N.~Pope,
``Higher-dimensional origin of D = 3 coset symmetries,''
  arXiv:hep-th/9909099.





\bibitem{Giusto:2007fx}
  S.~Giusto and A.~Saxena,
``Stationary axisymmetric solutions of five dimensional gravity,''
  Class.\ Quant.\ Grav.\  {\bf 24}, 4269 (2007)
  [arXiv:0705.4484 [hep-th]].

\bibitem{Gal'tsov:2008nz}
  D.~V.~Gal'tsov and N.~G.~Scherbluk,
``Generating technique for $U(1)^3 5D$ supergravity,''
  Phys.\ Rev.\  D {\bf 78}, 064033 (2008)
  [arXiv:0805.3924 [hep-th]].

\bibitem{Figueras:2009mc}
  P.~Figueras, E.~Jamsin, J.~V.~Rocha and A.~Virmani,
``Integrability of Five Dimensional Minimal Supergravity and Charged Rotating
Black Holes,''
  Class.\ Quant.\ Grav.\  {\bf 27}, 135011 (2010)
  [arXiv:0912.3199 [hep-th]].


\bibitem{Bossard:2008sw}
  G.~Bossard, H.~Nicolai and K.~S.~Stelle,
``Gravitational multi-NUT solitons, Komar masses and charges,''
  Gen.\ Rel.\ Grav.\  {\bf 41}, 1367 (2009)
  [arXiv:0809.5218 [hep-th]].

\bibitem{Harmark:2004ch}
  T.~Harmark and N.~A.~Obers,
``General definition of gravitational tension,''
  JHEP {\bf 0405}, 043 (2004)
  [arXiv:hep-th/0403103].


\bibitem{Myers:1999psa}
  R.~C.~Myers,
``Stress tensors and Casimir energies in the AdS/CFT correspondence,''
  Phys.\ Rev.\  D {\bf 60}, 046002 (1999)
  [arXiv:hep-th/9903203].


\bibitem{Bossard:2009at}
  G.~Bossard, H.~Nicolai and K.~S.~Stelle,
``Universal BPS structure of stationary supergravity solutions,''
  JHEP {\bf 0907}, 003 (2009)
  [arXiv:0902.4438 [hep-th]].


\bibitem{Copsey:2005se}
  K.~Copsey and G.~T.~Horowitz,
``The role of dipole charges in black hole thermodynamics,''
  Phys.\ Rev.\  D {\bf 73}, 024015 (2006)
  [arXiv:hep-th/0505278].


\bibitem{Gibbons:2004ai}
  G.~W.~Gibbons, M.~J.~Perry and C.~N.~Pope,
``The first law of thermodynamics for Kerr - anti-de Sitter black holes,''
  Class.\ Quant.\ Grav.\  {\bf 22}, 1503 (2005)
  [arXiv:hep-th/0408217].


\bibitem{Chong:2005hr}
  Z.~W.~Chong, M.~Cvetic, H.~Lu and C.~N.~Pope,
``General non-extremal rotating black holes in minimal five-dimensional
 gauged supergravity,''
  Phys.\ Rev.\ Lett.\  {\bf 95}, 161301 (2005)
  [arXiv:hep-th/0506029].


\bibitem{Shmakova:1996nz}
  M.~Shmakova,
``Calabi-Yau Black Holes,''
  Phys.\ Rev.\  D {\bf 56}, 540 (1997)
  [arXiv:hep-th/9612076].



\bibitem{Bena:2005ni}
  I.~Bena, P.~Kraus and N.~P.~Warner,
  ``Black rings in Taub-NUT,''
  Phys.\ Rev.\  D {\bf 72}, 084019 (2005)
  [arXiv:hep-th/0504142].



\bibitem{Gauntlett:2002nw}
  J.~P.~Gauntlett, J.~B.~Gutowski, C.~M.~Hull, S.~Pakis and H.~S.~Reall,
  ``All supersymmetric solutions of minimal supergravity in five dimensions,''
  Class.\ Quant.\ Grav.\  {\bf 20}, 4587 (2003)
  [arXiv:hep-th/0209114].




\bibitem{Bena:2004wv}
  I.~Bena,
``Splitting hairs of the three charge black hole,''
  Phys.\ Rev.\  D {\bf 70}, 105018 (2004)
  [arXiv:hep-th/0404073].




\bibitem{Emparan:2004wy}
  R.~Emparan,
 ``Rotating circular strings, and infinite non-uniqueness of black rings,''
  JHEP {\bf 0403}, 064 (2004)
  [arXiv:hep-th/0402149].


\bibitem{Maldacena:1997re}
  J.~M.~Maldacena,
``The large N limit of superconformal field theories and supergravity,''
  Adv.\ Theor.\ Math.\ Phys.\  {\bf 2}, 231 (1998)
  [Int.\ J.\ Theor.\ Phys.\  {\bf 38}, 1113 (1999)]
  [arXiv:hep-th/9711200].


\bibitem{Cvetic:1999ja}
  M.~Cvetic and F.~Larsen,
``Statistical entropy of four-dimensional rotating black holes from
near-horizon geometry,''
  Phys.\ Rev.\ Lett.\  {\bf 82}, 484 (1999)
  [arXiv:hep-th/9805146].


\bibitem{Emparan:2009cs}
  R.~Emparan, T.~Harmark, V.~Niarchos and N.~A.~Obers,
``Blackfolds,''
  Phys.\ Rev.\ Lett.\  {\bf 102}, 191301 (2009)
  [arXiv:0902.0427 [hep-th]].


\bibitem{Emparan:2009at}
  R.~Emparan, T.~Harmark, V.~Niarchos and N.~A.~Obers,
``Essentials of Blackfold Dynamics,''
  JHEP {\bf 1003}, 063 (2010)
  [arXiv:0910.1601 [hep-th]].



\bibitem{Emparan:2009vd}
  R.~Emparan, T.~Harmark, V.~Niarchos and N.~A.~Obers,
``New Horizons for Black Holes and Branes,''
  JHEP {\bf 1004}, 046 (2010)
  [arXiv:0912.2352 [hep-th]].


\bibitem{Emparan:2007wm}
  R.~Emparan, T.~Harmark, V.~Niarchos, N.~A.~Obers and M.~J.~Rodriguez,
``The Phase Structure of Higher-Dimensional Black Rings and Black Holes,''
  JHEP {\bf 0710}, 110 (2007)
  [arXiv:0708.2181 [hep-th]].



\bibitem{Kunduri:2006uh}
  H.~K.~Kunduri, J.~Lucietti and H.~S.~Reall,
``Do supersymmetric anti-de Sitter black rings exist?,''
  JHEP {\bf 0702}, 026 (2007)
  [arXiv:hep-th/0611351].

\bibitem{Kunduri:2007qy}
  H.~K.~Kunduri and J.~Lucietti,
``Near-horizon geometries of supersymmetric AdS(5) black holes,''
  JHEP {\bf 0712}, 015 (2007)
  [arXiv:0708.3695 [hep-th]].



\bibitem{Caldarelli:2008pz}
  M.~M.~Caldarelli, R.~Emparan and M.~J.~Rodriguez,
``Black Rings in (Anti)-deSitter space,''
  JHEP {\bf 0811}, 011 (2008)
  [arXiv:0806.1954 [hep-th]].





\end{thebibliography}
\end{document}